\documentstyle[prb,aps,preprint,tighten,epsfig]{revtex}
\begin{document}
\title {Weakly disordered spin ladders}

\author{E. Orignac and T. Giamarchi}
\address{Laboratoire de Physique des Solides, Universit{\'e} Paris--Sud,
                  B{\^a}t. 510, 91405 Orsay, France\cite{junk}}
\maketitle
\date{Rev 25/8/97 -- \today}
\begin{abstract}

We analyze an anisotropic spin 1/2 two legs ladder in the 
presence of various type of random perturbations. The generic
phase diagram for the pure system, in a way similar to 
spin one chains, consists of four phases:
an Antiferromagnet, a Haldane gap or a
Singlet phase (depending on the sign of interchain coupling) and two
XY phases designated by XY1 and XY2. The effects of disorder
depend crucially on whether it preserves 
XY symmetry (random field along $z$ and random exchange) or not
(random anisotropy or random $XY$ fields). In all cases we computed
the new phase diagram and the correlation length for the disordered system. 
The ladder exhibits a remarkable stability to disorder with XY symmetry.
Not only the singlet phases but also the {\bf massless} XY1 phase
are totally unaffected by disorder in stark contrast with the single
chain case. Even in the presence of an external magnetic 
field breaking the spin gap most of the phase diagram (including the XY point)
remains unaffected by the disorder, again in opposition with the 
single chain case. This stabilization towards disorder is similar,
albeit stronger, to the one occuring for fermionic ladders.
On the other hand the XY2 phase is very strongly
suppressed. Disorder breaking XY symmetry has much stronger
effects and most of the phases at the exception of the singlet
one are now destroyed. Interestingly, the disordered XY1 phase has 
a much shorter correlation length than the disordered XY2 phase 
contrarily with the case of perturbations preserving the rotation
symmetry around the z axis. The ladder 
system is thus in fact much more anisotropic than its single chain
counterpart with identical exchanges.
Finally we examine the case of strong disorder or weak interchain
exchange for XY symmetric disorder. Close to the isotropic point,
when the interchain exchange is 
increased the system undergoes a transition between two decoupled 
disordered spin 1/2 chains to a singlet ladder unaffected by disorder.
For more XY like systems, the transition occurs in two steps 
with a {\bf non disordered} XY1 phase occuring between the 
decoupled chains and the singlet phase. Comparison with 
other disordered gapped systems such as spin one chains is 
discussed.
\end{abstract}

\widetext
\section{Introduction}

One dimensional antiferromagnetic spin 1/2 systems are fascinating systems.
They can present quasi-long range order, without symmetry 
breaking yet with critical spin spin correlation functions and
diverging magnetic susceptibilities \cite{affleck_houches}. Not surprisingly, 
owing to the 
Jordan-Wigner transformation, such properties are remarkably
similar to the ones of interacting one dimensional fermions and are
known under the generic name of Luttinger liquid physics
\cite{solyom_revue_1d,emery_revue_1d}. Experimental realization of 
spin 1/2 chain provide a good confirmation of these 
remarkable properties\cite{tennant_kcuf3_expt,kim_srcuo2_expt}. In
addition, in a  way reminiscent of 
one dimensional fermions , spin chains are extremely sensitive to disorder.
In fact using the Jordan-Wigner
transformation\cite{jordan_transformation}, spin chains can 
be mapped onto  one dimensional spinless fermions, where the $J_z$ coupling gives 
rise
to an interaction term. Spin 1/2 systems are thus just another,
and maybe more convenient experimental realization of disordered
interacting fermions (or bosons)\cite{giamarchi_loc,doty_xxz}.
The transition to the disordered phase has been analyzed numerically 
in Refs. \onlinecite{runge_xxz,pang_dmrg,schmitteckert_dmrg} in good agreement 
with renormalization
group treatments \cite{giamarchi_loc,doty_xxz}.

Recently it was realized that these remarkable properties get drastically
modified when spin 1/2 chains are coupled together leading to the 
so called ladder system \cite{dagotto_2ch_review}.  In this case,
in a way very similar to the Haldane spin-S problem 
\cite{haldane_gap,schulz_spins}, a gap is found
to open for an even number of chains while the system remains massless
for an  odd number of chains. This phenomenon has been thoroughly investigated
both analytically \cite{strong_spinchains,gopalan_2ch,shelton_spin_ladders} and 
numerically \cite{white_2ch,%
sandvik_srcuo,hida_2ch,poilblanc_4ch,greven_2ch}.
In the recent years, progress in
solid-state chemistry resulted in the apparition of compounds with a
2-legs ladder structure such as SrCu$_2$O$_3$,
(La,Sr,Ca)$_{14}$Cu$_{24}$O$_{41}$ 
and Cu$_2$(1,4-Diazacycloheptane)$_2$Cl$_4$, in which the
theoretical predictions about the gap have received an eclatant confirmation
\cite{takano_spingap,chiari_cuhpcl,chaboussant_cuhpcl,%
hammar_cuhpcl,carter_lacasrcuo}.

Given the new physics present in these systems
it is  of paramount importance both from a theoretical
point of view and for practical experimental systems, to investigate
the effects of disorder in these systems.
This is especially true for the two leg ladder system for 
which zinc doping experiments are already performed \cite{azuma_zinc_doping}. 
Due to the formation of a gapped singlet
phase for the isotropic system it is somehow natural, although
not always trivial, to 
expect that weak disorder will have no effect. Most of the studies
about the influence of disorder have
been thus devoted so far to the effects of strong disorder on
such a gapped phase either in ladders or in other gapped systems
\cite{hyman_dimer_ranexchange,hyman_spin1_ranexchange,%
monthus_spin1_ranexchange,steiner_peierls_dirac,gogolin_disordered_ladder}
However, in the spin 1 chain system, the gap is proportional to the
intra chain exchange, whereas in the ladder system, the gap is
directly controlled by the interchain hopping $J_\perp$, and thus 
can be made small at will compared to the intra chain exchange 
$J$ permitting the use of perturbative methods to analyze the
competition of randomness and singlet formation.
Moreover, very little is known on the effects of disorder on the anisotropic 
system, and in particular in its gapless phases.
A surprising effect found for  fermionic two leg ladders 
\cite{orignac_2chain_short,orignac_2chain_long} in which a gap opens
in all the modes except the symmetric charge mode
\cite{nersesyan_2ch,schulz_2chains} 
is a remarkable stability 
of the ladder system to disorder. Whether a similar stabilization by gap
opening exists for 
two coupled spin chains needs to be investigated.  
We undertake such a study in the present paper. 
Quite remarkably we find that an anisotropic spin ladder
is completely stable {\bf even} in the gapless phase any
weak disorder respecting the XY symmetry. This is in stark contrast
with a single chain or even with ladders of spinless fermions. 

The plan of the paper is the following:
In section \ref{basics}, we  recall the basics of the bosonization
treatment of  spin chain systems. In order to study the effects of 
disorder on the anisotropic spin ladder we first investigate in details
in Section \ref{2chains-XY2} the phase diagram of the pure system. 
In addition to the three phases discussed in Ref. 
\onlinecite{strong_spinchains}, namely a Singlet phase, an XY1 phase and 
the antiferromagnetic phase, we find another XY phase, the XY2 phase,
in analogy with the case of spin one chains \cite{schulz_spins}. 
We discuss the possibility of observing this XY2 phase at 
intermediate coupling.
In section \ref{RZFields} we investigate the response of the two chain system
with respect to weak disorder preserving the rotational symmetry around the z 
axis, i.e. random fields parallel to the z axis and random exchange.
The spin ladder shows a remarkable stability to these perturbations 
even in the gapless phase. We also consider the disordered ladder
under a magnetic field, able to suppress the gapped phase. 
Finally we study the crossover between a single chain and the ladder 
behavior obtained by increasing the interchain coupling.
Given the remarkable stability of the ladder to random perturbations 
respecting the XY symmetry it is interesting to also 
investigate that breaks rotational symmetry
around the z axis. We thus study 
in section \ref{RXYFields} random anisotropy and random fields in the plane 
perpendicular to the z axis. Quite surprisingly the effect of
these perturbations is opposite to the one respecting the XY symmetry:
the XY1 gets very unstable, whereas the XY2 
phase is much less affected. Finally conclusions and open questions 
can be found in Section~\ref{sec:concl}.

\section{bosonization of one dimensional spin 1/2 chains}\label{basics}

In this section, the bosonization of
spin 1/2 chains is briefly recalled. For the sake of definiteness,  an
XXZ spin chain will be considered.
It is described by the following Hamiltonian:
\begin{equation}
\label{hamiltonien-XXZ}
H_{\text{XXZ}}=\frac{J_{xy}}{2}
\sum_{i}(S_{i+1}^{+}S_{i}^{-}+S_{i}^{+}S_{i+1}^{-})
+J_z\sum_{i}S_{i+1}^{z}S_{i}^{z}
\end{equation}
Spin chains Hamiltonians can be transformed into interacting 1
dimensional fermion systems  by expressing the spin operators
$S^+,S^-,S^z$ in terms of  fermion
operators $a^\dagger, a$ using the Jordan Wigner transformation
 \cite{affleck_houches,jordan_transformation,nijs_equivalence}
\begin {eqnarray} 
S_{i}^{+}=(-)^ia^{\dagger}_{i} \cos\left(\pi\sum_{j=0}^{i-1}a_{j}^{\dagger}
 a_{j}\right) \nonumber \\
S_{i}^{-}= (-)^i\cos\left(\pi\sum_{j=0}^{i-1}a_{j}^{\dagger} a_{j}\right)
a_i \label {Jordan-Wigner} \\
S_i^z=a^{\dagger}_{i}a_{i}-\frac 1 2 \nonumber
\end {eqnarray}
This changes the XXZ model into a  model of spinless fermions with
nearest neighbor interaction described by the Hamiltonian
\begin{equation}\label{spinless-fermions}
H=-J_{X}/2\sum_n( a^\dagger_n a_{n+1}+ a^\dagger_{n+1} a_n) +J_Z \sum_n
(a^\dagger_i a_i-\frac 1 2)(a^\dagger_{i+1} a_{i+1}-\frac 1 2)
\end{equation}
$J_z=0$ corresponds to free fermions and
the zero magnetization sector corresponds to an half-filled band
with Fermi-wavevector $k_F=\frac{\pi}{2a}$, $a$ being the lattice spacing.

To treat the problem with a finite $J_z$, it is convenient to use the
boson representation of one-dimensional fermion operators. Only
the final results are given here
since details about the method can be found in
various places in the literature
\cite{affleck_houches,solyom_revue_1d,emery_revue_1d,%
shankar_spinless_conductivite}. A detailed derivation with the same
notations can be found in the appendix~A
of Ref. \onlinecite{orignac_2chain_long}. The spin operators are given by
\begin{eqnarray} \label{bosonized-spin}
S^{+}(x) & = &\frac{e^{-\imath \theta(x)}}{\sqrt{2\pi a}}
\left[e^{-\imath \frac{\pi x} a}+\cos 2\phi(x) \right]\nonumber \\
S_z(x) & = & -\frac{1}{\pi}\partial_{x}\phi +
e^{\imath \frac{\pi x} a} \frac{ \cos 2\phi(x)}{\pi a}
\end{eqnarray}
Where $S^{+}(x)=\frac{S^{+}_n}{\sqrt{a}}$, $S^z(x)=\frac{S^z_n}{a}$ for
$x=na$, $a$ being the distance between two nearest neighbors sites
along the chain.
The phase $\phi$ is related to the average density of fermions (or
equivalently to the uniform spin density along $z$)
by $S_z(x) = -\frac{1}{\pi}\partial_{x}\phi$, whereas $\theta$ is
connected to the conjugate momentum $\Pi$ of $\phi$ (such that
$[\phi(x),\Pi(x')] = i \delta(x-x')$) by $\theta(x) = \int_{-\infty}^x
dy \Pi(y)$. In a very crude sense $\phi,\theta$ can be viewed as the
polar angles of a spin.
The Hamiltonian (\ref{hamiltonien-XXZ}) gives
\begin{equation}
\label{bosonized-XXZ}
H_{\text{XXZ}} = \int \frac{dx}{2\pi}\left[ u K (\pi \Pi)^2
+\frac{u}{K}(\partial_{x}\phi)^2
\right] +\frac{2\Delta}{(2\pi\alpha)^2}\int dx\cos(4\phi)
\end{equation}
This is a sine-Gordon Hamiltonian with $\beta=4$.
The cosine term comes from the so called
Umklapp process. Such processes are possible  as
$4k_F$ is equal to a reciprocal lattice vector
\cite{affleck_houches,nijs_equivalence,%
shankar_spinless_conductivite,luther_chaine_xxz}. The coupling constant can be 
determined
in perturbation in the interaction $J_z$ and gives for $J_z \ll J_{xy}$
\begin{eqnarray}\label{zorgl}
K &=& \left(1+\frac{4J_z}{\pi J_{xy}}\right)^{-1/2} \nonumber \\
u &=& aJ_X \left(1+\frac{4J_z}{\pi J_{xy}}\right)^{1/2}  \nonumber \\
\Delta &=& -J_z a
\end{eqnarray}
The range of validity of (\ref{bosonized-XXZ}) is much larger than the
simple perturbative regime provided the correct constant $K$ and $u$ are
used \cite{luther_chaine_xxz,haldane_xxzchain}. The isotropic point
$J_z = J_{xy}$ (Heisenberg model) corresponds in the bosonization
description to $K^*=\frac{1}{2}$,$\Delta^*=0$
as can be seen from the scaling dimensions of $S^+ $ and $S^z$
(see Eqs. (\ref{theta-theta}) and (\ref{phi-phi}) of
appendix~\ref{sinecor}). These correlation function become identical, as
they should for the Heisenberg point, only for
$K=\frac{1}{2}$. At that point
$\cos(4\phi)$ is marginally irrelevant. For larger values of $J_z$
($J_z > J_{xy}$) the cosine term becomes relevant. A gap opens in the
excitation spectrum of the $\phi$ field, leading to an insulating
charge density wave state for the fermions and equivalently to an Ising
order along $z$ for the spin chain (see e.g.
Ref.~\onlinecite{nijs_equivalence,shankar_spinless_conductivite,%
luther_chaine_xxz,haldane_xxzchain}).

Another instability of the XXZ model can be guessed from
(\ref{zorgl}). For $J_z \to -\frac{\pi}{4}J_{xy}$, $K \to \infty$.
This
is the sign of a transition to a fully polarized ferromagnetic state.
The exact solution shows that the ferromagnetic transition takes place for
$J_z=-J_{xy}$\cite{nijs_equivalence}. In the following, 
ferromagnetic systems  will not be considered
as their behavior is very similar to the one of
classical systems at $T=0K$.

\section{Two coupled spin 1/2 chains} \label{2chains-XY2}

In this section, two XXZ chains coupled by an exchange term are
considered using bosonization techniques.
The exchange coupling is of the form~:
\begin{equation}
\label{coupling-XXZ}
H_{\text{interchain}}=\sum_{i,\alpha=x,y,z} J_\perp^\alpha
S_{i,1}^\alpha S_{i,2}^\alpha
\end{equation}
where for simplicity, as for the intrachain Hamiltonian,
planar isotropy  $J_\perp^x = J_\perp^y = J_\perp^{x,y}$ is assumed.

The total Hamiltonian is~:
$H_{\text{pure}}=H_{\text{XXZ},1}+H_{\text{XXZ},2}+H_{\text{interchain}}$.
Each single chain Hamiltonian can be expressed in terms of fermions
operators using the Jordan-Wigner transformation like in
section~\ref{basics}. One has to take care that the ``fermions''
operators should have anticommutation relations inside one chain but
{\em commutation} relations between the chains to respect the spin
commutation relations. This causes no modification of the bosonized
expressions (\ref{bosonized-spin}) for the spin operators. Indeed it is
easy to check that (\ref{bosonized-spin}) preserves all the correct spin
commutation relations.
$H_{\text{XXZ}}$   has been obtained in section \ref{basics} and  only  the 
bosonized
expression of $H_{\text{interchain}}$ is needed.
Using (\ref{bosonized-spin}) and 
keeping only the most relevant operators one obtains
\cite{schulz_spins,strong_spinchains,luther_spin1}:
\begin{eqnarray} \label{exchange-bosonized}
H_{\text{interchain}}& = & \int \left[\frac{2g_1}{(2\pi a)^2}
\cos(\theta_1-\theta_2) +\frac{2g_2}{(2\pi a)^2}\cos 2(\phi_1-\phi_2)
 +\frac{2g_3}{(2\pi a)^2}\cos 2(\phi_1+\phi_2)\right] dx \nonumber \\
& & +J_\perp^z a \int dx \frac{\partial_x \phi_1 \partial_x
\phi_2}{\pi^2}
\end{eqnarray}
Where :
\begin{eqnarray}\label{g-function-J-perp}
g_1=\pi J_{\perp}^{x,y}a \nonumber\\
g_2=J_{\perp}^z a\nonumber\\
g_3=J_{\perp}^z a
\end{eqnarray}

The total Hamiltonian $H_{\text{pure}}$ is
rewritten  in terms of the fields
$\phi_{a}=\frac{\phi_{1}-\phi_{2}}{\sqrt{2}}$ and
$\phi_{s}=\frac{\phi_{1}+\phi_{2}}{\sqrt{2}}$ giving
\begin{eqnarray} 
H_{\text{pure}} & = & H_s +H_a \nonumber\\
H_s & = &\int \frac{dx}{2\pi}\left[ u_s K_s (\pi \Pi_s)^2
+\frac{u_s}{K_s}(\partial_{x}\phi_s)^2
\right] +\frac{2g_2}{(2\pi\alpha)^2}\int dx\cos(\sqrt{8}\phi_s)
\label{XXZ-final} \\
H_a & =&  \int \frac{dx}{2\pi}\left[ u_a K_a (\pi \Pi_a)^2
+\frac{u_a}{K_a}(\partial_{x}\phi_a)^2
\right] +\frac{2g_3}{(2\pi\alpha)^2}\int dx\cos(\sqrt{8}\phi_a)
+\frac{2g_1}{(2\pi\alpha)^2}\int dx\cos(\sqrt{2}\theta_a) 
\nonumber
\end{eqnarray}
Where :
\begin{eqnarray}\label{uKs-atsmallJperp}
u_s=u\left(1+\frac{KJ_\perp^z a}{2\pi u}\right), \qquad
K_s=K\left(1-\frac{KJ_\perp^z a}{2\pi u}\right)\nonumber \\
u_a=u\left(1-\frac{KJ_\perp^z a}{2\pi u}\right), \qquad
K_a=K\left(1+\frac{KJ_\perp^z a}{2\pi u}\right)
\end{eqnarray}
$u, K$ being given by (\ref{zorgl}) for a XXZ model and small $J_z$, and
more complicated expressions in the general
case \cite{luther_chaine_xxz,haldane_xxzchain}. If the correct values 
of $u$ and $K$ are used (\ref{XXZ-final}) is valid irrespective of
the values of $J_z$  and $J_{xy}$ and only assumes $J_\perp a
\ll \frac u K $. Because of the choice
of an hermitian string operator,  the
$\cos(4\phi_{1})$, $\cos(4\phi_{2})$ terms in $H_{\text{XXZ},1}$ and
$H_{\text{XXZ},2}$ respectively can be dropped,
since they always give terms that 
are less relevant than the terms that come from $H_{\text{interchain}}$.
Note that the interchain coupling in (\ref{XXZ-final})
is different from the one existing in
spinless fermions
ladders \cite{nersesyan_2ch,orignac_2chain_short,orignac_2chain_long}.
This is due to the presence of the string operator  (\ref{Jordan-Wigner}) 
in the fermion-spin equivalence.
Therefore two coupled spin chains have a different physics 
that cannot be directly deduced from the fermionic 
results\cite{orignac_2chain_short,orignac_2chain_long}. 
This distinction between spin chains and spinless fermions will
be crucial in the presence of disorder. 

Eqs. (\ref{exchange-bosonized}) and (\ref{XXZ-final})
have been derived, for particular cases or for physically related
systems, in various places in the literature.
Such Hamiltonians were first introduced to describe a quantum spin-one
chain \cite{schulz_spins,luther_spin1}. For this problem, although
the form of the Hamiltonian are identical $g_1$, $g_2$, $g_3$ are
related to parameters with a physical meaning different of the one of
$J_\perp$. For two coupled spin chains
(\ref{exchange-bosonized}) and (\ref{XXZ-final}) have been obtained and
investigated \cite{strong_spinchains,strong_spinchains_long}
for the particular case of isotropic interchain coupling 
($J^{x,y}_\perp=J^z_\perp$). 
This corresponds to
taking  $g_2=g_3=g_1/\pi$ in
(\ref{XXZ-final}). More recently, another
derivation for isotropically coupled spin chains 
has been obtained by a different method\cite{shelton_spin_ladders}. 
Instead of using
a Jordan-Wigner transformation, Ref.~\onlinecite{shelton_spin_ladders}
uses a representation in term of localized fermion. Such a representation is
particularly well adapted to describe an isotropic coupling since the
superexchange leading to the spin Hamiltonian is automatically
generated. Interestingly one recovers the same bosonized Hamiltonian
than (\ref{XXZ-final}) (with $K=1/2$ due to
isotropy), but with a different relation between the coefficients $g$. In
Ref.~\onlinecite{shelton_spin_ladders} the coefficient satisfy
$g_2=g_3=g_1\propto J_\perp$.
This difference comes from the fact that in the present 
derivation the calculation of the coefficients $g$ is perturbative and
the expression (\ref{g-function-J-perp}) is accurate for $J_z \ll J_{xy}$.
In going to the isotropic point these
coefficients get renormalized by irrelevant terms that do no affect the
form of the low energy Hamiltonian (\ref{XXZ-final}), but can change the
explicit value of the coefficients compared to
(\ref{g-function-J-perp}). Comparison of the two limits $J_z \ll J_{xy}$
and $J_z \sim J_{xy}$ shows that this renormalization effects are
rather weak, and that (\ref{XXZ-final}-\ref{uKs-atsmallJperp}) provide
an extremely accurate description 
of the physics of the ladder system for all values
of $J_z$, provided one does not need the {\em precise} values of the
coefficients $g$.

\begin{figure}
\centerline{\epsfig{file=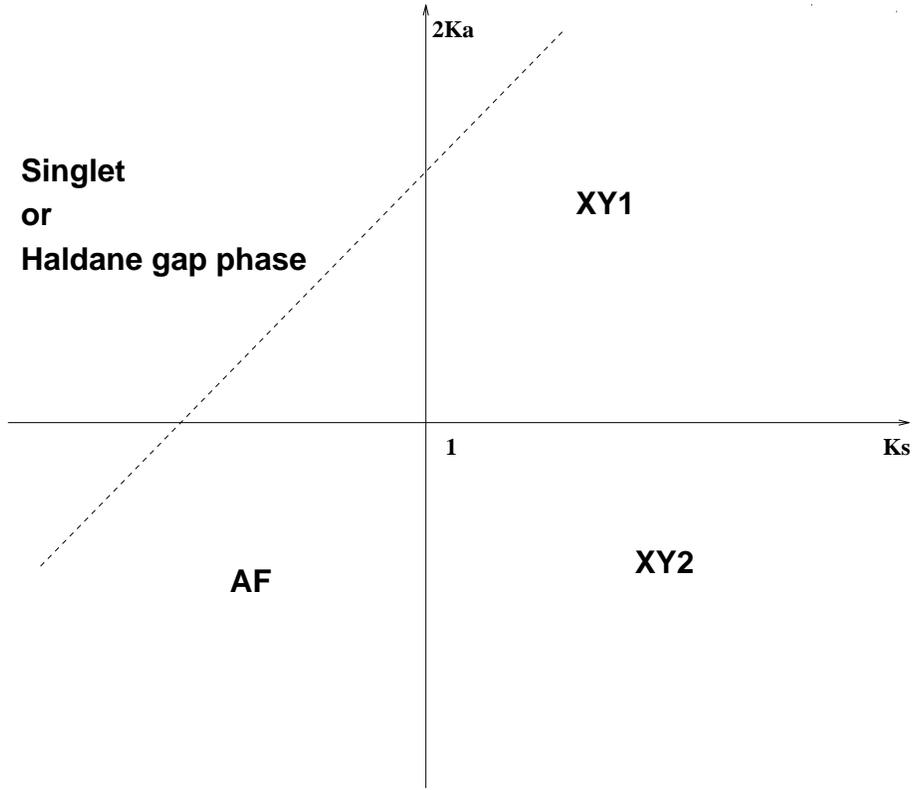,angle=-90,width=12cm}}
\vspace{0.5cm}
\caption{
The phase diagram of the pure 2 chain XXZ model in terms of
$K_a$ and $K_s$ XY1 and XY2 are gapless phases (see text). AF
contains antiferromagnetic quasi-long range order. The singlet 
(antiferromagnetic interchain coupling) or Haldane (ferro. interchain
coupling) have a gap to all excitations. The dotted line represent
the weak interchain coupling case, when the intrachain anisotropy
is varied. The isotropic point is $K_s=K_a=1/2$. The dashed line
corresponds to two coupled XXZ chains.
\label{pure-XXZ}}
\end{figure}

Using   (\ref{XXZ-final}) the phase diagram of the pure
system can be derived. 
Since the main goal of this paper is to analyze the additional 
effects of disorder, we restrict ourselves for the study of the pure system 
to isotropic interchain coupling but the intrachain coupling
is arbitrary. Nevertheless,
(\ref{XXZ-final}) can be used in  more general cases.
The derivation is technically similar 
to the one for the spin one chain \cite{schulz_spins}, but gives of course 
different physical phases. 
Since $H_s$ is a standard sine-Gordon Hamiltonian
(see appendix~\ref{sinecor})
its spectrum has a gap  if  $K_s<1$ and is gapless otherwise.
The treatment of $H_a$ is a bit more subtle.
$\cos(\sqrt{8}\phi_a)$ and $\cos(\sqrt{2}\theta_a)$ have respective
scaling dimension of $2K_a$ from
(\ref{phi-phi}) and $(2K_a)^{-1}$ from (\ref{theta-theta}) so
that both are relevant for $1/4<K_a<1$.
However, the coefficients of these operators are of the same order of
magnitude. Therefore, the most relevant operator is the first to
attain the strong coupling regime under renormalization. In the strong
coupling regime this operator takes a mean value that minimizes the
ground state energy. $\theta_a$ and $\phi_a$ being
conjugated, when one of them develops a mean value the exponentials
of the other one have zero expectation values and exponentially
decaying correlations (see appendix~\ref{sinecor}).
Therefore, $\phi_a$  acquires a
mean value while $\theta_a$ is disordered
for $K_a<1/2$  and the situation is reversed  for $K_a>1/2$.
As a consequence four different phases exists, similarly to the spin one case 
\cite{schulz_spins}, regardless of the sign of the interchain coupling.
Three of these four phases correspond to the one 
derived in Ref.~\onlinecite{strong_spinchains,strong_spinchains_long},
but, as discussed in detail below an additional massless phase can exist.
For isotropic intrachain coupling the results 
of Ref.\onlinecite{shelton_spin_ladders} are of course recovered. 
A summary of the phase diagram is shown on Figure~\ref{pure-XXZ}.
Let us now discuss each phase in details. 
The physics of these phases depends whether interchain 
coupling is ferro or antiferromagnetic, since this changes 
the average value taken by the massive fields.

\subsection{Ferromagnetic interchain coupling}\label{2chains-ferro}

A summary of the physics of each phase can be found in Table~\ref{table3}. 
A qualitative understanding of the results of table
\ref{table3}  can be obtained in the limit
$\mid J_\perp\mid \gg J,J_z$. Then, the ground state energy is minimized
by forming a spin one on each rung with the 2 spins 1/2. That case is
thus physically identical to the spin one
antiferromagnet\cite{schulz_spins}. The same physics hold even if
$J_\perp \ll J,J_z$. As shown in Table~\ref{table3} four sectors
exist.

In Sector I, the two spins 1/2 on every rung point in the same
direction parallel to the z axis while each chain
is antiferromagnetically ordered giving an
effective spin one Ising Antiferromagnet.

In sector II  a completely gapped phase exists
where all spin-spin correlation functions decay exponentially. Using
the analogy with the antiferromagnetic spin 1 chain, that phase is
identified with a Haldane gap phase \cite{haldane_gap}. Singlets are formed
here along the \emph{legs} of the ladder \cite{nijs_dof}.
The identification can be made more complete by
proving that the VBS order parameter, characteristic of a Haldane phase,
is effectively non-zero (see appendix~\ref{VBS-bosonization}).

Sector IV is a phase where only XY correlations remains.
This XY1 phase has order parameter $S_1^++S_2^+=S^+$ in agreement
with the physical picture of a total spin equal to one. Semiclassically, the XY1
phase tends to have the two spins in the XY plane. This phase has only a
gap in the antisymmetric spin excitations and has gapless
symmetric excitations. (See table \ref{table3}, sector IV for the
bosonized expression of the order parameter).

Finally, in sector III, $S_1^+S_2^+$ which
can be viewed as an effective raising operator for a spin 1/2 is an order 
parameter. Its bosonized expression
can be found in table~\ref{table3}. In the case of a true
spin 1 chain \cite{schulz_spins}, such a phase
can be obtained by adding an anisotropy of the form $-D(S_1^z+S_2^z)^2$
which favors the states  $S^z=+1,-1$ over $S^z=0$.
Thus, the XY2 phase is a phase in which the two spins on each rung can
only be found with either $(S^z_1,S^z_2)=(1/2,1/2)$ or
$(S^z_1,S^z_2)=(-1/2,-1/2)$. They can make transitions from one of these
state to the other and therefore
form an effective spin 1/2.  This accounts for the critical
fluctuations in that phase. The XY2 phase also shows subdominant Ising
Antiferromagnet fluctuations due to the existence of the intrachain
antiferromagnetic coupling. All other single spin fluctuations
decay exponentially fast.

At small $J_\perp$, (\ref{uKs-atsmallJperp}) gives
$K_s \sim K_a$. Thus, only the Haldane gap, XY1 and Ising Antiferromagnet
phase can be reached. These three phases were studied in Ref. 
\onlinecite{strong_spinchains,strong_spinchains_long}.
For isotropic coupling refermionization indeed gives an Haldane
phase \cite{shelton_spin_ladders}. When $J_\perp$ is very large the
effective Hamiltonian for the resulting spin one chain contains only
$J_{xy}, J_z$ terms, and thus only the same phases can be reached.
However, since when $J_\perp$ increases, (\ref{uKs-atsmallJperp}) indicate that
$K_a$ decreases, whereas $K_s$ increases it is interesting to
investigate whether one could reach the XY2 phase an intermediate
$J_\perp$. Of course answering this question probably needs a numerical
study of the model, since all the couplings would be of the same order.
Even if this interesting possibility does not occur, 
the XY2 phase can always be obtained by
using more complicated interchain couplings.

Finally, it is noteworthy that the only phase with  SU(2) symmetry
is the Haldane gap phase.
Two ferromagnetically coupled isotropic spin 1/2 chains, thus always
form a  Haldane gap phase, irrespective of the
strength of the coupling. The other phases are characteristic of
anisotropic systems.

\subsection{Antiferromagnetic interchain coupling}

Results for the antiferromagnetic coupling are summarized in
table~\ref{table2}. 

Sector I is again an Ising antiferromagnet phase. The spins on
the same rung point in opposite directions so that in lattice
representation, the order parameter
is $(-)^n (S_{n,1}^z-S_{n,2}^z)$. The boson representation of this
order parameter can be found in table~\ref{table2}.

Sector IV corresponds to the XY1 phase obtained
in  Ref.~\onlinecite{strong_spinchains}. Its  order parameter is
$S_1^+-S_2^+$. See table~\ref{table2} for the bosonized expression.
In that phase, the spins of each rung stay in
the XY plane and are antiparallel. They are free to rotate around the
z-axis provided they remain antiparallel and are antiferromagnetically
coupled to their neighbors giving rise to an XY phase.

Sector III is the other XY phase, the XY2 phase, for which
the order parameter is $S_1^+S_2^+$.
The hidden spin 1/2 degree of freedom corresponds now to the two states
of zero z component of the  spin formed by the addition of the spins
1/2 on the same rung. To the best of our our knowledge, such a phase was not
investigated in the previous
studies\cite{strong_spinchains_long,shelton_spin_ladders}.
The XY2 phase presents subdominant Ising Antiferromagnet fluctuations.
Thus, in the XY2 plane, there is a tendency to have the 2 spins 
parallel to the $z$ axis
and antiparallel among themselves. The intrachain coupling allows
transitions between these two states leading to critical fluctuations
analogous to the one of a spin 1/2 system.

Finally, there is a Singlet phase in which there is a gap to all
excitations and all correlation functions of spin operators decay
exponentially fast. A simple picture of that phase is obtained in the
limit $J_\perp/J_z \to \infty$.
In that limit,  the ground state can be seen  as made of singlets
along the rungs. In such a phase, there is an effective zero spin on
each site and accordingly all the spin-spin correlation function
are zero. For finite $J_\perp/J_z$, massive triplet excitations can propagate
along the chain leading to exponentially decaying spin-spin
correlation functions.

As in the case of a ferromagnetic interchain coupling, at small
$J_\perp/J_z$ the only possible phases are the XY1 phase, the
singlet phase and the Ising Antiferromagnet phase analyzed in
Ref.~\onlinecite{strong_spinchains}.
The XY2 phase appears at $K_s>1$ and $K_a<1/2$ (see table
\ref{table2}) and thus, as for the ferromagnetic case, cannot be
obtained for purely local intrachain coupling
and small $J_\perp$. Moreover, for
antiferromagnetic interchain coupling, increasing $J_\perp$ does not
favor the formation of the XY2 phase, so it is likely that one needs
here longer range interchain coupling. Finally, it is
important to remark that as in the case of the ferromagnetic
interchain coupling, the only phase that has  spin rotational
symmetry is the singlet phase, and that all other phases  can only exist
in systems with anisotropic interactions.

\section {Random fields and random planar exchange}\label{RZFields}

Since we now have  a complete description of the phases in the pure
system,  the effects of various type of disorder on these phases can
be considered .
Very different physical effects occur depending on whether the 
disorder respects or breaks the 
$U(1)$ symmetry (i. e. the rotational symmetry around the $z$-axis)
of the system). For the case of symmetric couplings in the XY plane
for the pure system the first type of disorder is the one most likely
to occur physically. We consider its effects in the present section.
As we discuss in more detail later perturbations that break 
the $U(1)$ symmetry are nevertheless also interesting to study and 
are considered in  section~\ref{RXYFields}.

\subsection{Coupling to disorder}\label{coupling-to-u1-disorder}

Two type of $U(1)$ conserving disorder can be investigated. 
The first one is the standard random field along the $z$ direction,
while the other is the random exchange. 
If a representation of the spins in terms of hard core bosons
was used  such perturbations would correspond to
random potentials or random hopping along the
chains for the associated disordered boson 
problem\cite{orignac_2chain_bosonic}. 

\subsubsection{Random field along $z$}

Let us consider first the coupling to a random magnetic field along the $z$ 
direction. This  field is assumed  weak enough not to destroy the gaps
of the pure system.
The lattice Hamiltonian for that problem is 
$H=H_{\text{pure}}+H_{\text{ZF}}$ with
\begin{equation}
\label{random-zfield}
H_{\text{ZF}}=\sum_{i}\left[h_{i,1}^z S_{i,1}^z + h_{i,2}^z S_{i,2}^z\right]
\end{equation}

For simplicity the disorder is taken Gaussian and uncorrelated  from
chain to chain and from site to site:
\begin{eqnarray}
\overline{h_{i,1}^z h_{i,2}^z}=0 \\ 
\overline{h_{i,n}^z
h_{j,n}^z}=D\delta_{i,j}
\end{eqnarray}
First, a  bosonized representation of (\ref{random-zfield}) has to be
obtained .
To do so, one goes to the continuum limit, decompose the $h^z$
fields in a slowly varying ($q \sim 0$) and an oscillating part
($q\sim 2k_F=\frac{\pi}{a}$), introduce the bosonized expression for
spin operators (\ref{bosonized-spin}), and keep only in the coupling
to disorder the slowly varying terms, in a way similar to
Ref.~\onlinecite{giamarchi_loc,doty_xxz}.  The final result in terms of the 
$\phi_1$
and $\phi_2$ fields is~:
\begin{eqnarray}
\label{boson-randomzf}
H_{\text{ZF}} & = & \int dx
\left[h_{1}^{2k_F}(x)\cos 2\phi_1(x) + h_{2}^{2k_F}(x)\cos 2\phi_2(x)\right]
\nonumber\\
& - & \int \frac{dx}{\pi} \left[h_1^{q\sim
0}(x) \partial_x \phi_1(x) +h_2^{q\sim 0}(x) \partial_x
\phi_2(x)\right] 
\end{eqnarray}
with
\begin{eqnarray}
\overline{h_i^{2k_F}(x)
h_j^{2k_F}(x')} &=& D_{2k_F}a\delta(x-x')\delta_{i,j} \nonumber \\
\overline{h_i^{q\sim 0}(x) h_j^{q\sim 0}(x')} &=& D_{q \sim 0} a \delta
(x-x') \delta_{i,j}
\end{eqnarray}
All other correlators are zero.
Introducing the symmetric and antisymmetric combinations
$h_s^{q\sim 0}(x)=h_1^{q\sim 0}(x)+h_2^{q\sim 0}(x)$, and
$h_a^{q\sim 0}(x)=h_1^{q\sim 0}(x)-h_2^{q\sim 0}(x)$, 
(\ref{boson-randomzf}) is rewritten in terms of the fields $\phi_a,\phi_s$ 
as~:
\begin{eqnarray} \label{eq:depart}
H_{\text{ZF}} & = & H_{\text{ZF}}^{q\sim
0}+H_{\text{ZF}}^{q\sim 2k_F}  \\
H_{\text{ZF}}^{q\sim 0} & = & -\int \frac{dx}{\pi\sqrt{2}}
\left[h_s^{q\sim 0}(x) \partial_x \phi_s(x) +h_a^{q\sim 0}(x)
\partial_x \phi_a(x)\right] \label{randomzfq=0} \\
H_{\text{ZF}}^{q\sim 2k_F} & = & \int dx
\left[h_{1}^{2k_F}(x)\cos \sqrt{2}(\phi_a+\phi_s)(x)+ + h_{2}^{2k_F}(x)
\cos \sqrt{2}(\phi_a-\phi_s)(x)\right]  \label{randomzfq=2kF} 
\end{eqnarray}
(\ref{eq:depart}) is the starting Hamiltonian for the discussion
of the effect of a random $z$ field in the framework of
bosonization.  (\ref{randomzfq=0}) represents the coupling of the
system with the $q\sim 0$ components of the random field, and
(\ref{randomzfq=2kF}) the coupling with the $2k_F$ components.

The 
$\partial_x\phi_s$ term in   (\ref{randomzfq=0}) has to be discussed
first. Two cases have to be distinguished~:
when there is no gap in $\phi_s$,  a transformation
\cite{giamarchi_loc} 
$\phi_s \to \phi_s +\int^x dx' h_s^{q\sim 0}(x')$, 
results in  a new Hamiltonian of the form   (\ref{eq:depart})
with $h_s^{q\sim 0}=0$, and identical correlators for
the disorder. Therefore,
the $h_s^{q\sim 0}$ terms cannot
affect the low energy spectrum of the system and only affect its
correlation functions in a trivial way.
On the other hand, if there is a gap in the $\phi_s$ degrees of
freedom, through refermionization the problem in the $\phi_s$ sector is
equivalent to a filled band of fermions with a gap to excitations in
the upper band  and a small slowly varying  random chemical
potential. The random potential  being smaller than the gap, it cannot
induce transitions in the upper band (i.e. the
system is incompressible) and thus cannot disorder the system.
Therefore, in both cases, at weak disorder strength the $h_s^{q \sim 0}$ is 
always irrelevant. A similar situation occurs for the
$\partial_x\phi_a$ term in   (\ref{randomzfq=0}).
When $\phi_a$ has a gap  the problem is obviously identical to the
problem of the gap in $\phi_s$.
When $\theta_a$ is massive, the refermionization procedure shows that
computing the response to the $\partial_x\phi_a$ term amount to
computing the superconducting response function of a band
insulator. Clearly, this has an exponential decay so that the
random field coupled to $\partial_x \phi_a$ term cannot disorder the system.
Thus all the $q\sim 0$ terms (\ref{randomzfq=0}) have no
important effect on the low energy response of the 2 chain system and
can be discarded. If the disorder is stronger however, the $q\sim 0$
part of the disorder will however be able to destroy the gaps of the 
pure system. This case will be discussed later.

\begin{figure}
\centerline{\epsfig{file=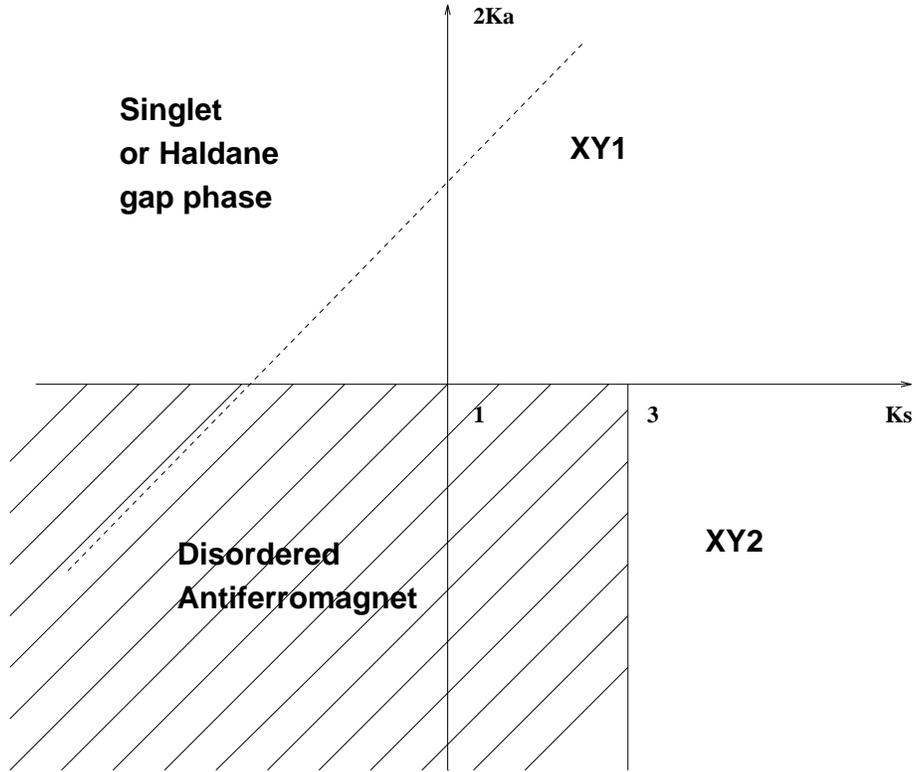,angle=-90,width=12cm}}
\vspace{0.5cm}
\caption{The phase diagram of the 2 chain XXZ model with a random $z$
field  in terms of $K_a$ and $K_s$. Lines indicate the parts of 
the phase diagram where disorder is relevant. The singlet phase is stable
due to the presence of the gap.
Quite surprisingly the massless XY1 phase is now also totally insensitive to 
weak disorder. The dashed line is for two coupled XXZ chains.
\label{RZF-XXZ}}
\end{figure}

Having shown that the $q\sim 0$ do not affect the phase diagram at
weak disorder, the $2k_F$ terms   (\ref{randomzfq=2kF}) remain to be
treated . 
To simplify the notations  the shorthand
$h_1$ is used for $h_1^{2k_F}$. The effect of the $2k_F$ disorder depends
on the phase that would exist in the pure system in the absence of disorder.
A detailed phase by phase discussion of the effects
of $2k_F$ randomness is given in section~\ref{sec:effet}.
The phase diagram in presence of disorder is shown
in figure~\ref{RZF-XXZ}. 

\subsubsection{Random exchange}\label{hamiltonian_for_random_exchange}

The other type of disorder that preserves the rotational invariance
around the $z$ axis is  random planar exchange and random $z$ exchange.
For convenience  the random $z$ exchange and random
planar exchange are treated separately.

Restricting to the small disorder case, the Hamiltonian in
the planar exchange case is $H=H_{\text{pure}}+H_{\text{PE}}$ with
\begin{equation}
\label{random-PE}
H_{\text{PE}}=\sum_{n} \left[
J_n^1(S_{n,1}^xS_{n+1,1}^x+S_{n,1}^yS_{n+1,1}^y)
+ J_n^2(S_{n,2}^xS_{n+1,2}^x+S_{n,2}^yS_{n+1,2}^y)\right]
\end{equation}
with $\overline{J_n^p J_{n'}^{p'}}=D\delta_{n,n'}\delta_{p,p'}$
$p=1,2$ being the chain index and $J_n^p \ll J$. The bosonized
Hamiltonian is given by \cite{doty_xxz}
\begin{eqnarray} \label{boson-randompe}
H_{\text{PE}} & = & \int dx
\left[J_1(x)^{2k_F} \sin2\phi_1(x) + J_2(x)^{2k_F}
\sin 2\phi_2(x) \right] \nonumber \\
              & + & \int dx \frac{v_F}{2\pi} J_1(x)^{q \sim 0}
\left[(\pi\Pi_1)^2 + (\partial_x\phi_1)^2 \right] \nonumber \\
             & + & \int dx \frac{v_F}{2\pi} J_2(x)^{q \sim 0}
            \left[(\pi\Pi_2)^2 + (\partial_x\phi_2)^2 \right]
\end{eqnarray}
where $J_p(x)$ is  the continuum limit of
$J_n^p$.

Let us now consider the random $z$ exchange case.
In the lattice spin 1/2 representation, the coupling to disorder is
represented by
\begin{equation}
H_{\text{random Jz}}=\sum_i \left[J_{i,1} S_{i,1}^zS_{i+1,1}^z +
J_{i,2} S_{i,2}^zS_{i+1,2}^z\right]
\end{equation}
The bosonized form for the $q\sim 0$ term is easily obtained in the
form
\begin{equation}
H_{\text{random } J^z }^{(q\sim 0)}=\int dx \left[J_z^{q\sim 0,1}(x)
(\partial_x \phi_1)^2(x) + J_z^{q\sim 0,2}(x) (\partial_x \phi_2)^2(x)
\right]
\end{equation}
Obtaining the $2k_F$ is not so straightforward. The derivation is
given in Appendix \ref{remplissage}. The final result is
\begin{equation}\label{randomJzq=2kf}
H_{\text{random} J^z}=\int dx \left[ J_z^{2k_F,1}(x)\frac{\sin
2\phi_1(x)}{2\pi a}+J_z^{2k_F,2}(x)\frac{\sin 2\phi_2(x)}{2\pi a} \right]
\end{equation}
with $\overline{J_z^{2k_F}(x)J_z^{2k_F}(x')}=D\delta(x-x')$.
The breakdown of bosonization for too large random planar
exchange can be read off in the $q\sim 0$ terms: if $q\sim 0$
disorder becomes too large the quadratic part of the Hamiltonian is no
more positive definite leading to a breakdown of the bosonization
description. However, if the weak disorder condition is met the
$q\sim 0$ terms couple to $\Pi^2$ and $(\partial_x\phi)^2$
terms. Power counting then implies that the $q\sim 0$ terms are 
irrelevant. Thus, the stability against weak random  exchange is again
determined only by the $2k_F$ terms.

\begin{figure}
\centerline{\epsfig{file=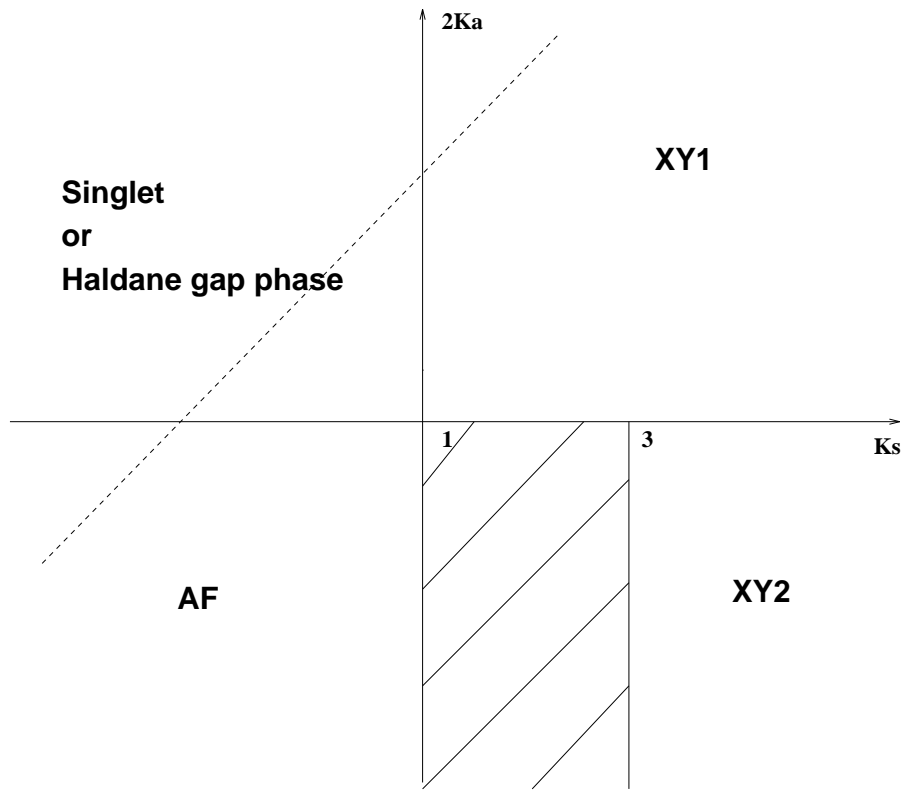,angle=-90,width=12cm}}
\vspace{0.5cm}
\caption{The phase diagram of the  2 chain XXZ model with a random
planar exchange or a random z exchange in terms of $K_a$ and $K_s$.
Lines indicated the parts of 
the phase diagram where disorder is relevant. Here again the XY1 
phase remains unaffected by weak disorder. The dashed line represents
coupled XXZ chains.
\label{RPE-XXZ}}
\end{figure}

Bosonization leads to the same Hamiltonians (\ref{boson-randompe}) and
(\ref{randomJzq=2kf}) for the coupling with the $2k_F$ part of random
planar exchange and random $z$ exchange. These two perturbations
can thus be treated the same way. In addition, 
in the phases in which $\phi_s$ is gapless, one can make the
transformation:
\begin{eqnarray} \label{transfo-gapless}
\phi_1 \to \frac \pi 4 +\phi_1 \nonumber\\
\phi_2 \to \phi_2 + \pi/4
\end{eqnarray}
changing the random  exchange  term into a random z field as can 
be seen by comparing (\ref{randomJzq=2kf}) and (\ref{boson-randompe})
for the random exchanges and (\ref{randomzfq=2kF}) for the random field.
Thus in the XY1 and XY2 phases both perturbations lead to the same 
stability regions and the same correlation lengths as is discussed in 
Section~\ref{sec:effet}. Of course the physical properties of the 
disordered phase are different depending whether the
disorder is random exchange or random field.
The phase diagram in the presence of random exchange is given in 
Figure~\ref{RPE-XXZ}.

\subsection{Effect of disorder} \label{sec:effet}

In this section  the effects of both the 
random $z$-field and the random exchange are considered in detail . 
The simpler case of the XY
phases is discussed first, then the case of the Haldane gap phase and
finally the case of the 
Antiferromagnet phase.

\subsubsection{XY2 phase}\label{xy2-with-u1-pert}

The XY2 phase 
i.~e.  The X-Y phase with non-zero mean value of $\phi_a$
(sector III of tables \ref{table3} and \ref{table2}) is considered.
For the sake of definiteness, only the random $z$ field case is
analyzed , 
but as discussed before in the XY phases 
the same results also apply in the random exchange
case. Using (\ref{randomzfq=2kF}) the effective 
Hamiltonian describing the coupling
with the impurity potential is 
\begin{eqnarray}
H_{\text{ZF}} &=& C \int dx (h_{1}(x)+h_{2}(x))\cos(\sqrt{2}\phi_{s})(x)
\qquad  \text{for } J_\perp <0  \label{rz-ferro} \\
H_{\text{ZF}} &=& C' \int dx(h_{1}(x)+h_{2}(x))\sin(\sqrt{2}\phi_{s})(x)
\qquad  \text{for } J_\perp >0  \label{rz-aferro} 
\end{eqnarray}
where $C=\langle \cos\phi_a \rangle$ and $C'=\langle \sin \phi_a
\rangle$ (see table~\ref{table3} and table~\ref{table2}).
The relevance of disorder can be determined by looking at the
renormalization of the disorder term \cite{giamarchi_loc}. 

The RG  equation for disorder is obtained from the scaling
dimensions of operators entering the coupling with disorder as
it is the same for both  (\ref{rz-ferro}) and (\ref{rz-aferro}).
\begin{equation}\label{bis-repetita} % placent
\frac {dD^{2k_F}} {dl} = (3-K_s)D^{2k_F}(l)
\end{equation}
Note that the RG equation for the disorder does not depend on the
nonuniversal constants $C$, $C'$.
The RG equation  for the Luttinger liquid
parameter $K_s$ is :
\begin{equation} \label{rg-for-K}
  \frac{dK_s}{dl}=-\Lambda^2 D
\end{equation}
where $\Lambda=C,C'$ depending on $\langle \phi_a \rangle$.
This equation is non-universal. However, for very weak disorder 
the region of stability is given by (\ref{bis-repetita}),
and one can discard (\ref{rg-for-K}) except to compute the critical
properties very close to the transition \cite{giamarchi_loc}.
Even for finite disorder (\ref{bis-repetita}) gives correctly 
the renormalized
value of the parameter $K$, and thus the exponents of the correlation
functions at the transition. 
 
Equation (\ref{bis-repetita}) immediately shows  that the XY2 phase is
unstable unless $K_s>3$.
It also gives the correlation length in the disordered phase, for weak
disorder and if one is not too close to the transition point
\cite{giamarchi_loc}.
The renormalization of $K_s$ at weak disorder,   (\ref{rg-for-K})
being negligible, the RG equation for $D$
can be integrated into
$D(l)=D(0)e^{(3-K_s)l}$.
This form of $D(l)$ is valid as long as $D\ll \frac {v_F^2} \alpha$,
since this condition
ensures that the energy scale induced by disorder is much smaller than
the energy cutoff. For $D(l) \sim \frac
{v_F^2} \alpha(0)$  disorder cannot be treated 
as a perturbation and  the RG has to be stopped . At that lengthscale,
the  system appears to be  strongly disordered.
Under the RG flow, the cut-off length has increased from $\alpha(0)$ to
$\alpha(l)=e^l \alpha(0)$ at which point the system
appears to be strongly disordered . Below it the system is described by
bosonization and disorder can be treated as a perturbation.
Thus  the length $\alpha(l)$ is the correlation length $\xi$, leading
to
\begin{equation} \label{corlength-randomzf}
\xi=\left(\frac{1}{D}\right)^{\frac{1}{3-K_s}}
\end{equation}
Using the transformation (\ref{transfo-gapless}), the same region
of instability $K_s<3$ and the same correlation length 
$\xi=(1/D)^{\frac{1}{3-K_s}}$ are obtained for the random exchange case.

\subsubsection{XY1 phase}\label{xy1-with-u1-pert}

 The case of the XY1 phase in the presence of a random
z field is now considered . This case is more involved since  to first order the
 effective coupling with disorder is zero (see tables \ref{table2},
\ref{table3}, sector IV). However, the coupling to disorder although irrelevant
by itself can generate relevant terms in higher order in perturbation.
Such terms can be computed, as was done in the case of the two
chain of fermions problem
\cite{orignac_2chain_short,orignac_2chain_long}, by integrating
over the massive modes. This leads to the effective Hamiltonian
for the coupling of the massless modes $\phi_s$ to disorder
\begin{equation}
\label{efctve-zfield}
H_{\text{ZF,eff.}}=\int dx \, \eta_{\text{eff}}(x)\cos(\sqrt{8}\phi_{s}(x))
\end{equation}
with $\overline{\eta_{\text{eff}}(x)\eta_{\text{eff}}(x')} \propto D^2
\delta(x-x')$.
By using the same method than above, the scaling dimension of 
the disorder is now $3 - 4 K_s$,
implying that the disorder  is  irrelevant unless $K_{s}<\frac{3}{4}$. But
at $K_{s}=1$
the $g_2$ term becomes relevant and drives the system towards
the singlet (if $g_2>0$) or the Haldane gap (if $g_2 <0$) phase
(see tables \ref{table2},\ref{table3}).
Thus, the XY1 phase is unaffected by weak random $z$-fields 
(or using the transformation (\ref{transfo-gapless}) by random exchange)
since these perturbations are irrelevant in its whole domain of existence.
The XY1 phase is thus unaffected by {\bf all} the perturbations respecting 
the rotation symmetry around the $z$ axis.

\subsubsection{The gapped phases}

 From section \ref{xy1-with-u1-pert} one sees easily that a two leg ladder 
system in a
singlet or a Haldane gap phase coupled to random $z$ fields is described
by the following effective Hamiltonian coming from integration of the
massive antisymmetric modes
\begin{equation}\label{singlet-disorder-z}
H=\int \frac{dx}{2\pi}\left[ u_s K_s(\pi\Pi_s)^2
+\frac{u_s}{K_s}(\partial_x \phi_s)^2\right] + \int dx \frac{2g_2
+\eta_{\text{eff.}}(x)}{(2\pi\alpha)^2} \cos \sqrt{8} \phi_s
\end{equation}
In the above discussion of the XY1 phase, it was shown that the non-random
$g_2$ term is more relevant
than the disorder term.
Thus  if the disorder is weak enough the deterministic
part of the $\cos \sqrt{8} \phi_s$ term
dominates over the random one and the gapped phases are stable with
respect to small random $z$-fields or small random exchange.
Such stability is reasonable on physical grounds in a singlet phase.
The stability to stronger disorder will be discussed in Section
\ref{sec:stab}.

\subsubsection{The Ising Antiferromagnet}

The Ising Antiferromagnet phase
(sector I in tables \ref{table3},\ref{table2}) is the simplest
to discuss due to its classical character. However, a marked
difference appears depending on whether one considers a random z field or
a random exchange.
For
$J_\perp <0$ one has
$\langle (S_1^z+S_1^z)\rangle \neq 0$ and for $J_\perp>0$,
$\langle (S_1^z-S_1^z)\rangle \neq 0$. For the random $z$-field,
a simple Imry-Ma type argument
\cite{shankar_spinless_conductivite} shows that the long range
order is lost and that the ground state of the
disordered 2 chain system is made of domains of characteristic size
$\sim 1/D^{2k_F}$.
This result is identical to the case of a classical
antiferromagnet. It can also be obtained directly on the spin 
Hamiltonian without using bosonization.
The Ising Antiferromagnet phase of the two leg ladder system is
thus unstable in the presence of  an arbitrarily
weak random z field, as was the case in the one chain problem
\cite{doty_xxz,shankar_spinless_conductivite}.
On the other hand, in the case of a weak random exchange, there is no
coupling at all to the disorder and the Ising Antiferromagnet is
\emph{stable} in the presence of disorder.

A summary of the phase diagrams for the random $z$ field and 
random exchange are respectively shown in Figure~\ref{RZF-XXZ}
and Figure~\ref{RPE-XXZ}.

\subsection{Discussion and Physical properties}

\subsubsection{Comparison with a single chain}

It is interesting to compare the above results with 
a single disordered spin chain. For a single chain only
two phases exist in the pure system: the Ising
antiferromagnet (for $K < 1/2$) and an XY phase (for $K > 1/2$)
that is the analogous of the XY1 phase for the ladder system.
The XY phase is destroyed by a random field along $z$ or random exchange
for $K < 3/2$ \cite{giamarchi_loc,doty_xxz,apel_spinless}.
The antiferromagnetic phase is unstable in the presence
of a random magnetic field but not random exchange as shown by an
Imry-Ma argument\cite{doty_xxz,shankar_spinless_conductivite}. 
The isotropic Heisenberg point is thus unstable to infinitesimal 
disorder. 

For the ladder case the isotropic point corresponds now to 
a singlet gapped phase. Quite naturally this phase is insensitive
to small disorder. The antiferromagnetic phase gives identical 
results than for a single chain. Quite remarkable however
the XY1 {\bf massless} phase is now resistant to all perturbations
respecting around the $z$ axis, at the opposite of the corresponding
phase for the single chain. This surprising result can be
understood by noticing that the XY1 phase is much more anisotropic 
than its one chain counterpart (in particular although it is massless
it still has {\bf exponentially} decaying $2k_F$ correlations of 
the $S_z$ component). The disorder can only couple to higher 
operators that are less relevant, in a way reminiscent of the 
situation in the fermion ladder problem 
\cite{orignac_2chain_short,orignac_2chain_long}.

The opposite situation occurs for the XY2 phase. 
This phase phase reveals itself much more unstable than the XY phase
in the presence of random $z$ fields or random exchange. Its domain of
existence is reduced to $K_s>3$. As a result, it correlation length
$l^{XY2}_{2ch.}\sim \left(\frac 1 D
\right)^{\frac 1 {3-K_s}}$ is also much shorter than its 
single chain counterpart $l_{\text{1 ch.}}\sim \left(\frac 1
D\right)^{\frac 1 {3-2K}}$. This is due to the existence of strong
antiferromagnetic fluctuations in the XY2 phase that easily couple to disorder.
It is again reminiscent of the easy coupling of the charge density
wave phase for the fermionic ladder 
\cite{orignac_2chain_short,orignac_2chain_long}.
The two most remarkable effects occurring in the ladder system
are thus the stabilization of the isotropic point due to the 
singlet phase and quite unexpectedly the stability of the 
XY1 massless phase as well. This remarkable stability against 
disorder prompt for several questions. In particular it is interesting
to understand how the phase diagram evolves if the gaps allowing 
for this stability against disorder are destroyed, either 
by a magnetic field or by increasingly strong disorder.
These two cases are examined in the next two sections. 

\subsubsection{Effect of a magnetic field}\label{uniform+random-z-field}

The coupling to a uniform magnetic field has the form:
\begin{equation}
H_{\text{uniform}}=\frac{h\sqrt{2}}{\pi}\int dx \partial_x\phi_s
\end{equation}
Therefore, a strong enough magnetic field, suppresses the gap
formation in $\phi_s$ but does not affect the 
antisymmetric sector. 
Thus the antiferromagnetic phase and the Haldane gap phase
disappear upon application of a strong enough magnetic field and 
are replaced by incommensurate phases\cite{chitra_spinchains_field}.
The stability of the remaining two incommensurate gapless XY phases in the 
presence of the random magnetic field or the random exchange can be analyzed
by similar techniques than the ones of section \ref{sec:effet}.

\begin{figure}
\centerline{\epsfig{file=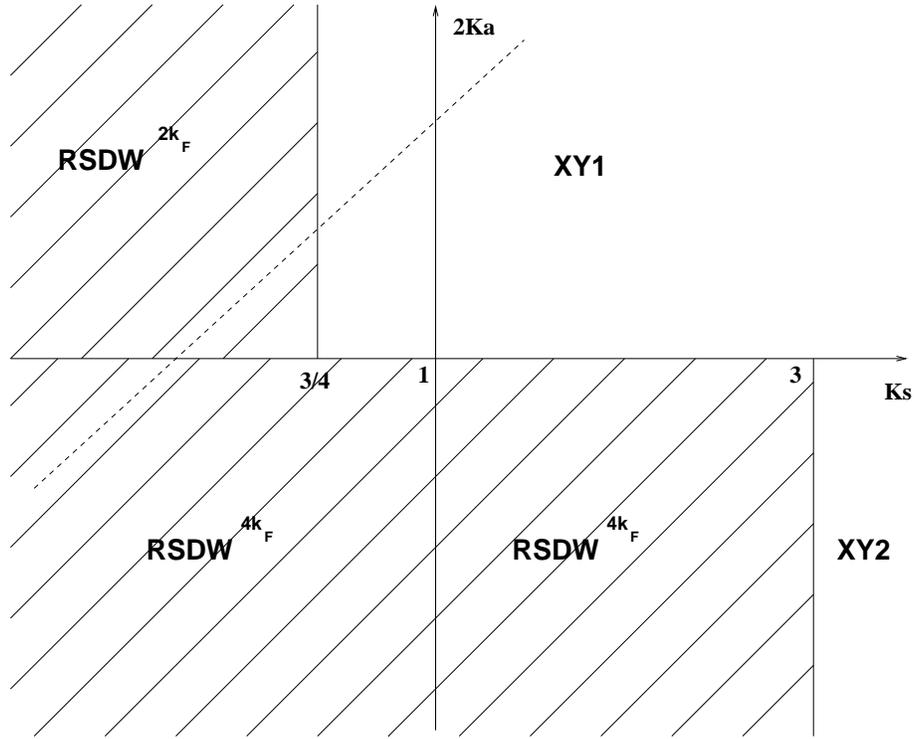,angle=-90,width=12cm}}
\vspace{0.5cm}
\caption{The phase diagram of the two chain XXZ model with a random 
z magnetic field under a uniform magnetic field parallel to the z axis.
The uniform magnetic field is strong enough to inhibit Haldane gap or
singlet gap
formation. This results in the apparition of a disordered phase.
The dashed line corresponds to  coupled XXZ chains. \label{fig:magnetic-field}}
\end{figure}

Using the analysis of  Sec. \ref{xy1-with-u1-pert},
one finds that the random $z$ field 
results in the suppression of the XY1 phase for
$K_s<3/4$ and formation of a random antiferromagnet with a
correlation length $l_{AF,4k_F}\sim \left(\frac 1 D \right)^{\frac 2
{3-4K_s}}$. The modulation of the spin density is at 
a wavevector $4k_F= 2\pi (1+2m)$ , where $m$ is the 
magnetization and is incommensurate
\cite{chitra_spinchains_field} with the the lattice spacing.
Contrarily to the zero magnetic field case, the disordered $4k_F$ 
antiferromagnet is no more wiped out by Haldane gap formation.
According to  Sec. \ref{xy2-with-u1-pert}, a random $z$ field 
also results in suppression of the XY2 phase for $K_s<3$ and
formation of a random $2k_F$ antiferromagnet . The situation in this
case is however identical to the zero magnetic field case.
The results are summarized on figure \ref{fig:magnetic-field}.
In particular, for $1/2<K_s<3/4$ the system is again sensitive to disorder.
The case of a random exchange is similar except that the $4k_F$
antiferromagnet is replaced by a $4k_F$ random singlets phase.

Using the standard mapping of spins on hard core bosons, the 
problem of the coupled chains under field can be related 
to the more general problem of a bosonic ladder.
This problem will be analyzed in details elsewhere 
\cite{orignac_2chain_bosonic}.
In the boson language, the XY phases correspond to superfluid phases,
and antiferromagnetic phases to charge density waves. Random z fields
correspond to random potentials and random planar exchange to random hopping.

\subsubsection{Stability to strong disorder} \label{sec:stab}

It is also interesting to look on how the gaps in the system
can be destroyed by the disorder itself, when it becomes strong enough.
Strictly speaking such a study is beyond the reach of a renormalization
group treatment since such a transition would correspond to 
a transition between two strong coupling fixed points
, in a regime where the RG is not 
rigorously applicable. Fortunately one can still make 
some physical arguments. It would be thus of great practical
interest to check whether the simple analysis performed here
can be confirmed in more sophisticated treatments such as 
simulations or by using non-perturbative methods 
\cite{dasgupta_rg,fisher_random_transverse,fisher_rtfim,%
gogolin_disordered_ladder,orignac_mott_variat}.

For simplicity we concentrate on the Haldane (or singlet) phases 
and on the XY1 phase in the presence
of a random magnetic field. Both the symmetric and antisymmetric gap,
when they exist can be destroyed by either the $q \sim 0$ or $q \sim
2k_F$
component of disorder.
A similar effect has been analyzed 
in the case of non-magnetic impurities in
Ref. \onlinecite{fujimoto_2ch_halffilling_disorder}  using
bosonization and RG techniques.
The $q\sim 0$ component of the magnetic field
can prevent $\phi_a$ from developing a gap, 
however if $\theta_a$ develops a gap, 
(i. e. $K_a>1/2$) this component
of the random magnetic field has simply no effect  
for the reasons exposed in Sec. \ref{coupling-to-u1-disorder}.

Three regimes of stability can thus a priori be defined. At very small 
$J_\perp$ one has disordered decoupled spin $1/2$ chains where no 
gap exists both in the symmetric and antisymmetric sector. The correlation
length in this regime is $\xi_{\text{decoupled}} \sim (1/D)^{1/(3-2K)}$, using 
the approximation $K_s \sim K_a \sim K$ for small $J_\perp$. This phase 
is unstable when the correlation length due to the opening of one 
of the gaps by $J_\perp$ becomes comparable to $\xi_{\text{decoupled}}$. 
It is easy to check that the shortest correlation length induced by $J_\perp$
is the antisymmetric one.
The transition occurs therefore when 
\begin{equation}
J_\perp^* \sim D^{(2-1/2K)/(3-2K)}
\end{equation}
thus giving $J_\perp^* \sim D^{1/2}$ for $K=1/2$ (the isotropic point) and 
$J_\perp^* \sim D^{3/2}$ for $K=1$ (the limit to the XY1 phase).
For larger values of $J_\perp$ a gap exists in the antisymmetric 
$\theta_a$ mode. 

\begin{figure}
\centerline{\epsfig{file=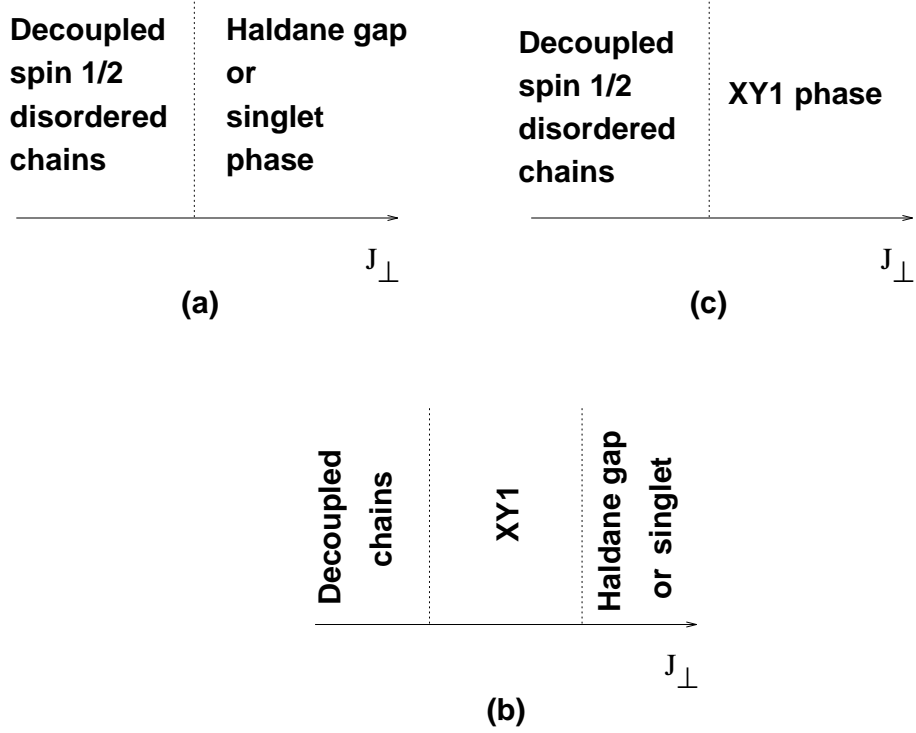,angle=-90,width=12cm}}
\vspace{0.5cm}
\caption{Phase diagram for fixed disorder as a function of the 
strength of the interchain coupling $J_\perp$. (a) For $K=1/2$
the isotropic point. There is a direct transition between decoupled
disordered spin $1/2$ chains and a stable Haldane or singlet phase.
(b) For $K \ge 0.76465$ a disordered XY1 phase exists between
the decoupled chains phase and the stable Haldane phase.
This phase shows {\bf no } topological order. 
(c) For $K >1$ A direct transition between decoupled chains 
and the stable XY1 phase takes place. 
\label{increasej}}
\end{figure}

Since the gap in the symmetric mode behaves as $J_\perp^{1/(2-2K)}$,
the properties of this mode crucially depends on the value of $K$. 
The critical value of $J_\perp$ needed to resist the $2k_F$ and the $q\sim 0$
 of the  random $z$ field  
of the disorder 
are respectively
\begin{eqnarray} \label{critical}
J_{\perp,\text{c1}} \sim D_{q\sim 0}^{2-2K_s} \\
J_{\perp,\text{c2}} \sim D_{q\sim 2k_F}^{\frac{4-4K_s}{3-4K_s}} 
\end{eqnarray}
For the isotropic point $K=1/2$ both these values are smaller than
$J_\perp^*$.  By increasing $J_\perp$ one should thus go directly
from two disordered decoupled spin $1/2$ chains to a stable 
Haldane (or singlet) ladder system. 
Although one can only give physical arguments for the nature of the 
transition, the crossing of two correlation length suggest that there is
a  sudden drop of the gap to zero. 
It is clear that the decoupled chains do not have any form of topological
hidden order and that their response function is simply the one
of isolated disordered chains. The phase diagram is shown on 
Figure~\ref{increasej}-a.

As one make the system more anisotropic, and moves towards the 
XY1 phase the gap in the symmetric sector decreases. The critical 
values of $J_{\perp,c1}$  become larger than 
$J_\perp^*$ for $K\ge 0.76465$. For such values of $K$ ,the $2k_F$
component of the random z-field is irrelevant. 
Therefore, for
$J_\perp^*<J_\perp<J_{\perp,\text{c1}}$ a gap is generated by
interchain coupling in the antisymmetric mode but not in the
symmetric mode due to the $q\sim 0$ component of the random z field.
In other words, an intermediate gapless XY1 phase 
exists between the decoupled chain phase and the stable Haldane phase.
Since the VBS order parameter is the correlation function $\langle\cos 
\sqrt{2}\phi_s(x)
\cos \sqrt{2}\phi_s(x')\rangle$ in the limit $\mid x-x' \mid$ (see appendix 
\ref{VBS-bosonization}) it is strictly zero in the phase
that results from the breaking of the Haldane gap,
due to the presence of the forward scattering disorder.
This situation is shown in Figure~\ref{increasej}-b.
If $K >1$, $J_\perp$ is irrelevant in the symmetric 
sector and one recovers a direct transition between the decoupled 
chains and the stable XY1 phase, with algebraic correlations,
as shown on Figure~\ref{increasej}-c.

For random exchange, it has been shown in Sec. 
\ref{hamiltonian_for_random_exchange} 
that the $q\sim 0$ component of disorder is irrelevant in the RG
sense.
 For  values of $K$ in the interval $[ 0.76465,1]$ the 
$q\sim 2k_f$ part of the disorder is irrelevant, so random exchange
which does not give rise to a relevant $q\sim 0$ term will continue
to give the transitions of Figure~\ref{increasej}-a. For $K>1$, random
exchange disorder gives the transitions of Figure~\ref{increasej}-c. 

It is noteworthy that the resulting phase diagrams for disorder
larger than the gaps in the ladder system is quite  different from the one 
where {\bf no} coupling of the antisymmetric mode to the 
disorder could occur such as for a true spin one chain. 
In that case increasing the disorder would lead to a disordered 
Haldane phase that could retain some degree of VBS order 
for the random exchange 
case \cite{hyman_spin1_ranexchange,monthus_spin1_ranexchange}. 

\section{Random perturbation breaking the U(1) symmetry} \label{RXYFields}

 From a theoretical point of view it is also interesting to consider
randomness that break the $U(1)$
rotational symmetry of the XXZ model. Indeed for the single chain
case \cite{doty_xxz},
such disorder was proved to be very efficient in destroying the
quasi long range order in the system. Given the remarkable stability of
the gapless phase of the ladder system to the type of disorder examined
in section~\ref{RZFields}, it is interesting to check whether the same
property still occurs.  In this section, the two most common
types of disorder breaking the $U(1)$ symmetry are considered, namely a random 
field
confined to the $XY$ plane and random planar anisotropy.

\subsection{Coupling to disorder}

\subsubsection{Random field in the XY plane}

The  two chain system with a random magnetic field
 in the XY plane is considered.
The Hamiltonian  is $H=H_{\text{pure}}+H_{\text{XYF}}$,
$H_{\text{XYF}}$ being given by

\begin{equation}
\label{randomXYfield}
H_{\text{XYF}}=\sum_i \left[ h_{i,1}^xS_{i,1}^x+ h_{i,1}^yS_{i,1}^y +
h_{i,2}^xS_{i,2}^x+ h_{i,2}^yS_{i,2}^y \right]
\end{equation}
With $\overline{h_{i,p}^a h_{j,q}^b}=D\delta_{i,j}\delta_{p,q}\delta_{a,b}$
($a,b=x,y$ $p,q=1,2$).

To bosonize this expression, it is convenient to rewrite (\ref{randomXYfield})
in terms of $S_{i,a}^{\pm}$, $(a=1,2)$ and introduce
$h_{i,a}^{\pm}=h_{i,a}^x\pm\imath h_{i,a}^y$.
This gives
\begin{equation}
\label{randomXYfield+-}
H_{\text{XYF}}=\frac{1}{2}\sum_i \left[ h_{i,1}^+S_{i,1}^-+
h_{i,1}^-S_{i,1}^+ +
h_{i,2}^+S_{i,2}^- + h_{i,2}^-S_{i,2}^+ \right]
\end{equation}
For chain 1, upon bosonization the coupling with the random XY field
can be rewritten as\cite{doty_xxz}:
\begin{equation}
H_{\text{XYF,1}}=\int \frac{dx}{\sqrt{8 \pi}a} \left[h_1^{(q \sim 0)+}
e^{-\imath \theta_1(x)}{\cos 2\phi_1(x)}
+ h_1^{(2k_F)+}e^{-\imath \theta_1(x)}\right] +\text{H. c.}
\end{equation}
A similar expression holds for chain 2.
 From power counting, the most relevant
terms in the bosonized Hamiltonian are the $2k_F$ ones.
One could be tempted to keep only the $2k_F$ terms and simply drop
the $q\sim 0$ ones.
However, the $q\sim 0$ part of the coupling to disorder has a bosonized form
$e^{\pm\imath \theta}\cos(2\phi)$ as can be seen from (\ref{bosonized-spin}) 
and generates in second order in perturbation theory
a term $\int dx h_{\text{eff},1}^z(x)\cos(2\phi_1)(x)$ i.e an
effective random magnetic field parallel to the $z$ axis
(see   (\ref{bosonized-spin})) with
$h_{\text{eff.},1}^z(x)\propto h_1^{(q \sim
0)+}h_1^{(2k_F)-}+\text{h. c.}$.
That term is always  more relevant
than the  $q\sim 0$ term,
and can be relevant even when the $2k_F$ term is irrelevant.
It is easily seen that no other relevant terms are  generated so that
one has to 
keep just the generated  random $z$ field and drop the $q\sim 0$ term from the
Hamiltonian. This situation is typical of
random perturbations that break the
rotation symmetry around the $z$ axis \cite{doty_xxz}.
The generated random $z$-field transforms the random field restricted 
to the XY plane into a more general anisotropically distributed random field.
As a byproduct, the effect of an isotropic random field can also be
easily obtained. It is obvious that the phase boundaries induced by
the random XY field and the isotropic random field will be identical.
However, the correlation length will not show the
same dependence on disorder strength in the phases the properties of which
are dominated by the $z$ component of the random field.

 From the preceding discussion of the generated terms,
the bosonized Hamiltonian
containing only the most relevant terms is
$H=H_{\text{pure}}+H_{\text{XYF, bosonized}}$ with $H_{\text{pure}}$
given by (\ref{XXZ-final}) and 
\begin{eqnarray}
\label{XYField-boson}
H_{\text{XYF, bosonized}} & = &\int dx \left[
\frac{h^{(2k_F)-}_1(x)}{\sqrt{2\pi a}}
e^{\imath\frac{\theta_s+\theta_a}{\sqrt{2}}}
+\frac{h^{(2k_F)-}_2(x)}{\sqrt{2\pi
a}}e^{\imath\frac{\theta_s-\theta_a}{\sqrt{2}}} +
\text{H. c.} \right]\nonumber \\
   &  & +\int dx \left[
h_{\text{eff.},1}^z(x)\cos(\sqrt{2}(\phi_{a}+\phi_{s}))+
h_{\text{eff.},2}^z(x)\cos(\sqrt{2}(\phi_{a}-\phi_{s})) \right]
\end{eqnarray}
where $\overline{h_{\text{eff.},p}(x)h_{\text{eff.},p'}(x'})=C D^2 \delta(x-x')
\delta_{p,p'}$
The second line of the above equation is the random $z$-field
already studied in Section~\ref{RZFields}, the first line 
is the random planar field itself.

One can expect
the phases showing strong XY fluctuations to be more strongly affected
by the random XY field than the phases
having dominant antiferromagnetic fluctuations.
Indeed the latter ones only couple to the generated random $z$-field, 
or in other words their coupling to disorder
involves intermediate states with energies above the gaps.
Technically, this means that it is necessary to separate the case in which
$\phi_a$ is massive (i. e. dominant antiferromagnetic fluctuations) 
and the case in which $\theta_a$ is massive (i. e. dominant XY fluctuations).
Obviously, the singlet phase and the Haldane gap phase should be insensitive
to small random fields since these phases are completely gapped and 
have vanishing spin-spin correlations.

\subsubsection{Random planar anisotropy}

Another form of perturbation breaking the rotational symmetry around
the z axis  is when $\delta J=J_X-J_Y$ fluctuates randomly from site
to site with a mean value zero.
The relevant Hamiltonian  is  $H_{\text{pure}}+H_{\text{RA}}$ in which
\begin{equation}
\label{randomXYaniso}
H_{\text{RA}}=\sum_i \left[ \delta
J_i^1(S_{i,1}^xS_{i+1,1}^x-S_{i,1}^yS_{i+1,1}^y)+
\delta J_i^2(S_{i,2}^xS_{i+1,2}^x-S_{i,2}^yS_{i+1,2}^y)\right]
\end{equation}
With $\overline{\delta J_{i}^p \delta J_{j}^q}=D\delta_{i,j}\delta_{p,q}$
($p,q=1,2$). As in the case of the random XY fields,
contractions of the $q\sim 0$
and $q\sim 2k_F$ terms  generate relevant terms \cite{doty_xxz}.
The bosonized form of the $q\sim 0$ part is~:
\begin{equation}
\label{boson-RA}
H_{\text{RA}}^{q\sim 0}=\int dx \gamma_1(x)
\cos(\sqrt{2}(\theta_a+\theta_s))(x)
+\int dx \gamma_2(x)\cos(\sqrt{2}(\theta_s-\theta_a))(x)
\end{equation}
with $\gamma_p(x=na) \propto \delta J_n^p$,
and  the bosonized form of the $2k_F$ part is~:
\begin{eqnarray}
H_{\text{RA}}^{2k_F}& = &\int dx \gamma_1^{2k_F}(x)
\cos(\sqrt{2}(\theta_a+\theta_s))(x)
\cos(\sqrt{2}(\phi_a+\phi_s))(x)\nonumber \\
                     & + & \int dx
\gamma_2^{2k_F}(x)\cos(\sqrt{2}(\theta_s-\theta_a))(x)
\cos(\sqrt{2}(\phi_a-\phi_s))(x)
\end{eqnarray}
An effective random z field
is generated by contraction of the  $q\sim 0$ with the $2k_F$
part as in the case of the random field in the XY plane.
A
Its expression is
\begin{equation}\label{RZF-forRA}
H_{\text{generated}}=\int dx \left[ h_1^z(x)\cos(\sqrt{2}(\phi_{a}+\phi_{s}))+
h_2^z(x)\cos(\sqrt{2}(\phi_{a}-\phi_{s})) \right]
\end{equation}

where $\overline{h_{p}^z(x)h_{p'}^z(x')}=C D^2 \delta(x-x')
\delta_{p,p'}$
This random z field (\ref{RZF-forRA}) is always more relevant than the $2k_F$ 
part.
Therefore only (\ref{boson-RA}) and (\ref{RZF-forRA})
have to be kept in the bosonized Hamiltonian.

\subsection{Physical Properties}

Having worked out the bosonized representations of the perturbations 
that break the $U(1)$ symmetry, it is possible to analyze their
effects on the phases of the system.

\subsubsection{XY1 phase}

This phase is expected to be the most unstable in the presence of a
random XY field or a random anisotropy. Since the XY1 phase
is not affected by a random $z$-field (see Section~\ref{RZFields}),
the corresponding term can be safely dropped.  
$\theta_a$ is replaced by its average value 
$\langle \theta_a \rangle$ in (\ref{XYField-boson}) giving
the following simplified expression of the coupling
to the random XY field:
\begin{equation}\label{XYF-Bs1}
H_{\text{XYF, eff.}}\sim \int dx \left[\xi_{\text{eff.}}(x)
e^{\imath\frac{\theta_s(x)}{\sqrt{2}}}+\text{H. c.} \right]
\end{equation}
Where $\xi_{\text{eff.}}(x)=(\xi_1(x)+\xi_2(x))$ for $J_\perp<0$ and
$\xi_{\text{eff.}}(x)=\imath (\xi_1(x)-\xi_2(x))$ for $J_\perp >0$  
(see tables \ref{table3},\ref{table2}).

In the case of the random anisotropy, the effective coupling in the
XY1 phase has the form:
\begin{equation}\label{rpa-theta=0}
H_{\text{RA}}=\int \gamma_{\text{eff.}}(x)\sin(\sqrt{2}\theta_s(x)) dx
\end{equation}
for $J_\perp <0$ and:
\begin{equation}\label{rpa-thetane0}
H_{\text{RA}}=\int \gamma_{\text{eff.}}(x)\cos(\sqrt{2}\theta_s(x)) dx
\end{equation}
for $J_\perp >0$. $\phi_s$ being massless,  the RG equation for the
random XY field is derived from (\ref{XYF-Bs1})
in the form
\begin{equation} \label{eq:rgran}
\frac{dD}{dl}=\left(3-\frac{1}{4K_s}\right)D
\end{equation}
Equation (\ref{eq:rgran}) implies that the random XY field
is relevant for $K_s>1/12$. Since the XY1 phase only exists at $K_s>1$
(see tables \ref{table3},\ref{table2}) this phase is completely 
suppressed by an arbitrarily weak random magnetic
field in the XY plane. At small disorder the
correlation length of the disordered XY1 phase is 
\begin{equation} \label{corrlengthXY}
l_{\text{r. field}} \sim
(1/D)^{\frac{4K_s}{12K_s-1}}
\end{equation}

Similarly for the case of planar anisotropy, the RG equation
for the disorder is 
\begin{equation}
\frac{dD}{dl}=\left(3-\frac{1}{K_s}\right)D
\end{equation}
The disorder is thus relevant for $K_s>1/3$ and the XY1 
phase is also unstable to weak random anisotropy.
The correlation length is in that case
\begin{equation} \label{xy1-rpa-cl}
l_{\text{r. anisotropy}} \sim (1/D)^{\frac{K_s}{3K_s-1}}
\end{equation}
For the isotropic point $l_{\text{r. anisotropy}} \sim 1/D$.

\subsubsection{Singlet and Haldane gap phases}

In these two phases, the coupling with disorder has the same
simplified expression (\ref{rpa-theta=0}-\ref{rpa-thetane0}) and 
(\ref{XYF-Bs1})
than in the XY1 phase due to the identical
structure of the gaps in the antisymmetric modes. Moreover, it is  known
from section \ref{RZFields} that a random $z$-field has no effect.

However, the presence of a mass term in the symmetric part of
the Hamiltonian leads to the suppression of all disorder correlation
functions. The gapped phase is thus stable as for the case  of
perturbations preserving the rotational symmetry.

\subsubsection{The XY2 phase}

This phase has weaker XY fluctuations than the XY1 phase and should therefore
be less affected by the random XY field or the random anisotropy .
Nevertheless, it has subdominant antiferromagnetic fluctuations  
and can be disordered by coupling to the
generated random $z$ field.

Indeed it is known that random z fields  suppress  the
XY2 phase if $K_s<3$ (see Section \ref{RZFields}).
For $K_s>3$ on the other hand, only the random XY field or the random
anisotropy can possibly disorder the XY2 phase.
In the XY2 phase, $\phi_a$ is massive, and
this gives apparently zero coupling to disorder when simplifying
(\ref{XYField-boson}). 
In fact, an effective coupling of the
random XY field to $\phi_s$ can be derived through second order
perturbation theory along the lines of
Ref. \onlinecite{orignac_2chain_short,orignac_2chain_long}.
The calculations are  straightforward  and lead to  the following  effective
coupling to the random XY field
\begin{equation}
H_{\text{XYF, eff.}}=\int dx
\left[\xi_{\text{eff.}}(x)e^{\imath\sqrt{2}\theta_s}+ \text{H. c.}\right]
\end{equation}
where $\overline{\xi_{\text{eff.}}(x)\xi_{\text{eff.}}(x')}\propto D^2
\delta(x-x')$, and prefactors coming from mean values of the $\phi_a$
fields have been omitted.
A similar calculation for the random anisotropy case leads to the
following effective coupling:
\begin{equation}\label{RA-phi}
H_{\text{RA, eff.}}=\int dx \gamma_{\text{eff.}}(x) \cos(\sqrt{8}\theta_s(x))
\end{equation}
With $\overline{\gamma_{\text{eff.}}(x)\gamma_{\text{eff.}}(x')}
\propto D^2 \delta(x-x')$. \\

 From power counting, the RG equation for the random field is
\begin{equation}
\frac{dD_{XY}}{dl}=\left(\frac 3 2-\frac{1}{2 K_s}\right)D_{XY}
\end{equation}
implying that
the $2k_F$ part of the random XY field  is relevant as soon as $K_s >1/3$.
The correlation length in the disordered phase is
\begin{equation} \label{clxy2rfxy}
l_{\text{random field}} \sim (1/D_{XY})^{\frac{2K_s}{3K_s-1}}
\end{equation}

For random anisotropy, the RG equation is:
\begin{equation}\label{rg-ra}
\frac{dD_{RA}}{dl}=\left(\frac 3 2 -\frac{2}{K_s}\right)D_{RA}
\end{equation}
Since the random planar anisotropy is
relevant  for
$K_s>\frac{4}{3}$, the XY2 phase does not survive
the presence of a small random planar anisotropy also for $K_s>3$.
the correlation length of the resulting  disordered phase is  then given
by $l_{\text{random anisotropy}} \sim (1/D_{RA})^{\frac{2K_s}{3K_s-4}}$.

\subsubsection{The Ising antiferromagnet}

Since the antiferromagnetic phase is unstable
with respect to a weak random z-field (see section \ref{RZFields}), it
will be completely suppressed here by  the generated random $z$
field both in the case of random anisotropy and random field in the XY
plane. The corresponding disordered phase
is a random antiferromagnet of correlation length
$l\sim 1/D_z$ . This correlation length is just the one of the
disordered XY2 phase ``frozen'' at $K=1$.

\begin{figure}
 \centerline{\epsfig{file=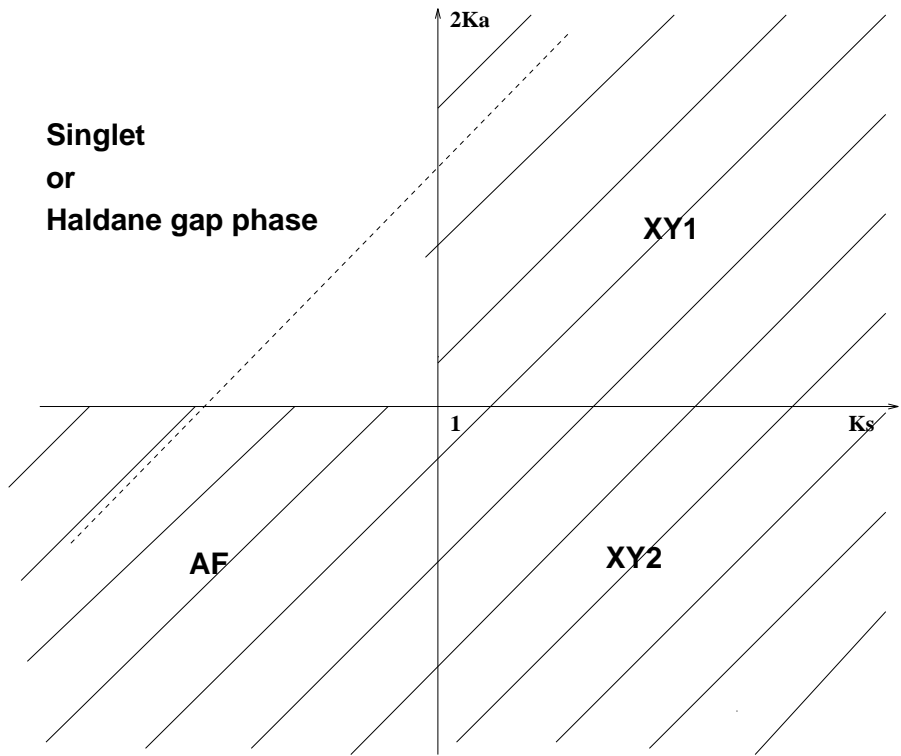,angle=-90,width=12cm}}
   \vspace{0.5cm}
\caption{The phase diagram of the  2 chain XXZ model with a random
field in the XY plane or a random planar anisotropy  in terms of $K_a$
and $K_s$. The dashed line corresponds to coupled XXZ chains.}
\label{RXYF-XXZ}
\end{figure}

The full phase diagram is given on figure  \ref{RXYF-XXZ}.

\subsection{Physical discussion}
\subsubsection{comparison with the one chain system}

In order to compare the two XXZ spin chain system  with a single XXZ chain, the 
discussion has to be
restricted  to the antiferromagnetic, singlet and XY1 phases since the
XY2 phase does not obtain in a two XXZ chain system.
The single chain system has a XY phase for $J_z<J_{XY}$ and an
antiferromagnetic phase for $J_z>J_{XY}$.
The XY phase of the single chain
is suppressed by a random field in the XY plane and the correlation length
is \cite{doty_xxz} $l_{\text{1ch.}} \sim
(1/D_{XY})^{\frac{2K}{6K-1}}$.
The XY1 phase in the two chain system is also suppressed, but :
$l_{\text{RF,2ch.}} \ll l_{\text{RF,1ch.}}$ (see   (\ref{corrlengthXY})).
Similarly, it is also unstable in the
presence of random anisotropy\cite{doty_xxz}, with a correlation length
$l_{\text{RA,1 ch.}} \sim (1/D_{RA})^{\frac{K_s}{3K_s-2}}$ in the
disordered phase. Again this length 
is much larger than its ladder counterpart 
$l_{\text{RA,2ch.}}\ll l_{\text{RA,1ch.}}$ (see   (\ref{xy1-rpa-cl})).
Thus, interchain coupling
makes an XXZ ladder systems much more sensitive to random
perturbations that break rotational symmetry around the z axis than
a single chain. In both case the effect of disorder is considerably
enhanced since the \emph{exponents} of the correlation length are 
changed.  
By contrast, for perturbations that preserve the rotation symmetry
around the z axis the one chain system was \emph{more} disordered than
its two chain counterpart. This seem to indicate that the XY phase in 
the ladder system
is in effect much more anisotropic than its single chain counterpart.

As for perturbations respecting the rotation symmetry, effects of 
disorder in the antiferromagnetic phase 
are quite similar for a single chain and for the ladder.
Indeed for a single chain, contrarily to the results of 
Ref. \onlinecite{doty_xxz} , the generated random z field 
suppresses the antiferromagnet phase, both in the case of a random field
in the XY plane and of random anisotropy.  
The same effect also holds in the two chain system. In that case, the
correlation lengths show the same dependence with disorder strength up
to prefactors.
Contrarily to the one chain case, there is no phase boundary between
the disordered XY phase and the random antiferromagnet phase due to
the existence of the Haldane  gap (or the singlet) phase which  is not
suppressed by a weak
disorder.

The XY2 phase is also suppressed. For a random field,
however, the correlation length in the disordered phase
is  much longer than the one of the disordered XY phase of a single
chain, $l_{\text{corr.}} \sim
(1/D)^{\frac{2K}{6K -1}}$.
In the presence of random anisotropy, the XY2 phase is also suppressed.
However, the XY2 phase gives rise to two different disordered phases.
The first one corresponds to suppression of the quasi long range order
by the random anisotropy per se, the other one to suppression of
QLRO by the effective random z field.
A crossover occurs between these two phases at the point where they have
identical correlation lengths.
The correlation length of the phase pinned on the random z field
being  given by $l_{\text{corr.}} \sim (1/D_z)^{\frac{2}{3-K_s}}$
where $D_z$ measures the strength of the effective random magnetic field. 
Since $D_z\sim D^2$ the crossover
should occur for $K_s=2$. for $K_s<2$, the properties are dominated by the
generated random z field and for $K_s>2$ by the random anisotropy.
Two different disordered phases already
existed in the XXZ chain with random planar anisotropy\cite{doty_xxz}.
Such an effect in the two chain system is specific of the XY2
phase and does not exists in the XY1 phase.
The correlation length of the disordered phase induced by a random
anisotropy in a XY2 phase is larger than
the correlation length of its one chain counterpart 
$l_{\text{corr.,1 ch. }} \sim (1/D)^{\frac{K}{3K_s-2}}$.
Paradoxically, although the XY2 phase does not exist 
in an XXZ two chain system with weak antiferromagnetic coupling, 
its properties {\bf regarding disorder} 
are much more similar to the one of the XY phase of the single XXZ
chain than the ones of the XY1 phase. Such difference is explained by
the fact that in the XY1 phase the $2k_F$ fluctuations in $S^z$ are
suppressed by gap formation leading to an increased robustness with
respect to perturbations that respect rotational symmetry around the z
axis and by the locking of the spins that sit on the same rung of the
ladder leading to an increased sensitivity to perturbations that break
rotation symmetry around the z axis. Also, it can be shown that in the
XY2 phase, the spins on each rung of the ladder can only take
identical values (for $J_\perp <0$) or opposite values (for $J_\perp
>0$) leading to an effective spin 1/2 chain.

\subsubsection{Effect of a uniform magnetic field}

\begin{figure}
 \centerline{\epsfig{file=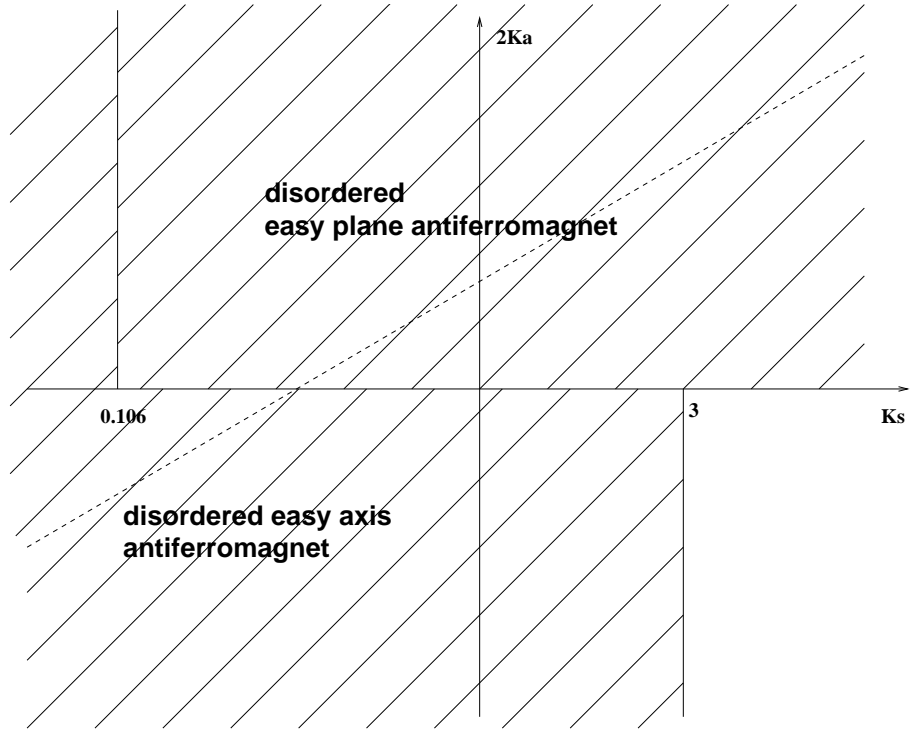,angle=-90,width=12cm}}
   \vspace{0.5cm}
\caption{The phase diagram of the two coupled spin chains in the presence of a 
random field
in the XY plane and a uniform field parallel to the z axis.
The dashed line corresponds to two coupled XXZ chains. The uniform
field inhibits Haldane gap formation.}

\label{fig:uniform+xy}
\end{figure}

\begin{figure}
 \centerline{\epsfig{file=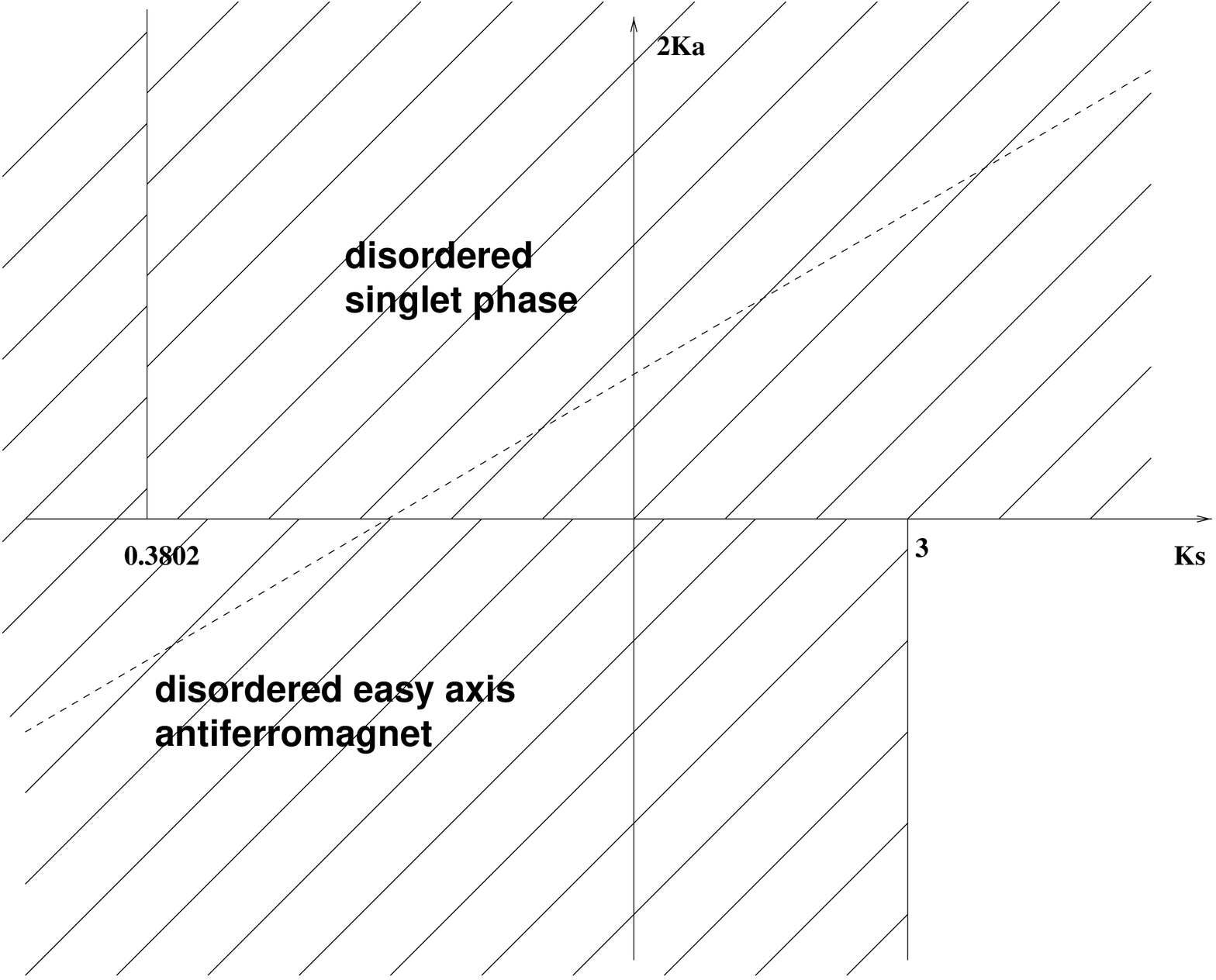,angle=-90,width=12cm}}
   \vspace{0.5cm}
\caption{The phase diagram of the two coupled spin chains in the presence of a 
random planar anisotropy and a uniform magnetic field parallel to the
z axis preventing Haldane gap phase formation. The dashed line
corresponds to two XXZ coupled chains.
}
\label{fig:uniform+pa}
\end{figure}

As seen in section \ref{uniform+random-z-field}, a uniform magnetic field
can inhibit Haldane gap formation.
In the presence of a random XY field or random anisotropy, this lead to the 
formation
of a disordered XY phase, with a correlation length  given by Eq.
(\ref{corrlengthXY})
for a random XY magnetic field and by Eq. (\ref{xy1-rpa-cl}) for a random
planar anisotropy.
It can be verified that the correlation length induced by the effective random 
z field is much longer than the correlation length induced by 
the random XY field for $K_s>\frac{\sqrt{97}-9}{8} \simeq 0.106$ 
and much longer than the correlation length induced by the random XY anisotropy
for $K_s>\frac{\sqrt{145}-9}{8}\simeq 0.3802$.
In other words, the induced random z field is a very weak perturbation
that takes over only when the random field in the XY plane or the random 
anisotropy is
nearly irrelevant.
Therefore, in the presence of an applied uniform magnetic field, a disordered
phase replaces the Haldane gap phase.
The resulting phase diagram can be found on figure \ref{fig:uniform+xy} for
the random magnetic field in the XY plane and on figure \ref{fig:uniform+pa} 
for the random planar anisotropy.

\section{Conclusion} \label{sec:concl}

In this paper we have investigated the effects of disorder on an 
anisotropic two leg spin ladder. 

For the pure system, we have computed the phase diagram in details.
Both for ferromagnetic and antiferromagnetic 
interchain coupling the system can exhibit four different phases:
an antiferromagnetically ordered state, a gapped singlet phase (Haldane phase)
and two massless XY phases. One of the XY phases (XY1) is a close 
analogous to the XY phase of a single chain. The other one (XY2), that 
was not discussed in previous work on the 2 coupled chain
system, is the analogous to the one occurring for a spin one chain 
with on site anisotropy. 
That phase has a tendency to have the spins parallel to the $z$
axis. For weak interchain coupling only the AF, Haldane and XY1 phases
can be realized. An interesting question is whether for intermediate coupling
one can stabilize or not the XY2 phase. Of course such a phase can 
always be realized by using more complicated interchain couplings.

The disordered ladder shows remarkable features compared to a single
spin chain. For perturbations respecting the XY symmetry such 
as a random $z$-field and random exchange, both the singlet phase
and the {\bf massless} XY1 phases revealed to be totally insensitive to 
weak disorder. On the other hand the XY2 phase 
is extremely sensitive to disorder due to the presence of 
strong antiferromagnetic fluctuations in the $z$ direction.
Similarly to the one chain case, the effect of randomness on
the Ising antiferromagnet depends on whether the perturbation is invariant under
$S^z \to -S^z$ or not. In the former case it does not affect the Ising
AF in the latter it suppresses it through an Imry Ma mechanism.
While such a stability was to be expected from the 
gapped singlet phase, it is much more surprising in the
massless XY phase and is reminiscent of the delocalization 
occurring in fermionic ladders. Note that here the effect is even 
stronger, since the delocalization transition only occurred for weakly
attractive interactions close to the non-interacting point. 
This would correspond to a non-disordered phase for $J_z<0$ and a
disordered phase for $J_z>0$ in the spin language, whereas the spin
system is stable with respect to disorder for $J_z<J$. 
In the presence of a finite magnetic field this stability to disorder
subsides, even if the singlet phase is destroyed. The isotropic point $K=1/2$
becomes now sensitive to disorder, but for moderate XY anisotropy
$K >3/4$ disorder has again no effect. This problem offers interesting 
connections to disordered bosonic ladders or coupled vortex planes.
This extreme stability to disorder for the spin ladder is to be 
strongly contrasted with the one for a single chain where most of 
the phase diagram is destroyed by infinitesimal disorder and only
extremely anisotropic XY chains $K >3/2$ can resist. 

Knowledge of the phase diagram for weak disorder allows to make 
reasonable guess on the behavior of the system for stronger disorder
or when the gaps are reduced (e.g. by diminishing the interchain 
couplings). Close to the isotropic point $K\sim 1/2$, upon increasing 
the interchain coupling, there should be 
a single transition between a regime of decoupled (and thus disordered)
spin 1/2 chains towards a stable singlet phase ladder for 
$J_\perp^* \sim D^{(2-1/2K)/(3-2K)}$. 
However by making the system more XY anisotropic
the transition should occur in two step. Decoupled spin chains would have 
a transition towards coupled chains, where the antisymmetric mode is 
gapped, but the symmetric mode is not gapped, leading to an XY1 
phase. Upon increasing the interchain coupling further the symmetric mode
gaps giving back the singlet phase. Thus for anisotropic ladders 
with $0.76465 < K < 1$  an intermediate {\bf non-disordered}
XY1 phase should appear for intermediate interchain couplings. 
For large anisotropies the singlet phase does not exist any more 
and one recovers a direct transition between decoupled chains and 
the stable XY1 phase of the ladder. Since this transition are obtained 
by the crude comparison of the various correlation length in the 
system it would be interesting to confirm such a phase diagram
by numerical simulations. Subtle effects might indeed occur for 
disorder such as the random exchange where the true correlation 
length of the system (compared to the one given by the RG around the 
Gaussian fixed point) is found to diverge \cite{fisher_rtfim}. 

The extreme stability of the ladder to XY symmetric randomness made 
it worth to investigate perturbations breaking this symmetry as well.
Remarkably the behavior is here inverted. Randomness 
breaking the rotational symmetry 
around the z axis, suppresses the two XY phases and the
Ising AF phase as was the case for one chain. However
the disordered  XY1 phase has a \emph{much shorter}
correlation length than the disordered XY phase of the one chain
system in the presence of the same perturbations whereas the XY2 phase
has a \emph{much longer} correlation length than the XY phase of
the one chain system. Such an effect is again due to the fact that
the ladder system with a given $J_z/J \neq 1$ is in effect much more
anisotropic than its 
one chain counterpart with the same $J_z/J$ parameter. 
The XY2 phase in the presence of  planar anisotropy 
gives rise to two different disordered phases similarly
to the XY phase of the one chain system but at odds with
the XY1 phase that gives rise to only one phase. Such result
is at first sight paradoxical since the XY1 appears as the natural 
continuation
of the XY phase of the one chain XXZ system, whereas the XY2 phase
is likely to exist only in more complicated models or at larger
coupling.

In the presence of a strong enough uniform magnetic field, the Haldane
gap phase is suppressed, allowing for the observation of a crossover
from a phase dominated by the random field in the XY plane or the
random planar anisotropy to a phase dominated by a generated random
field parallel to the z axis. This crossover occurs in the vicinity of
the transition to the disordered classical antiferromagnet and may
therefore be difficult to probe in experiments or numerical simulations.

Clearly the disordered ladder system presents an extremely rich behavior
and the renormalization study presented here can only be a first step towards
its understanding. This is specially true for the crossover to strong 
disorder, which would be interesting to investigate through numerical
simulations and  non perturbative methods. 

\acknowledgments

We are grateful to R. Bhatt, R. Chithra, P. Le Doussal, A. J. Millis,  
and H.J. Schulz for useful 
discussions. E. O. thanks ISI for hospitality and support for participating the
1996 Euroconference of  ``The role of Dimensionality in the Correlated
Electron Systems'' held in Torino May 6-25, 1996.
T. G. thanks the ITP where part of this work was 
completed for support under NSF grant PHY94-07194.
 
\appendix

\section{sine-Gordon Hamiltonians} \label{sinecor}

The general  sine-Gordon Hamiltonian has the form:
\begin{equation}
\label{basic-SG}
H_{SG}=\int \frac{dx}{2\pi}
\left[uK(\pi\Pi)^2+\frac{u}{K}(\partial_x\phi)^{2}\right]
+\Delta \int dx \cos(\beta \phi)
\end{equation}
To get some understanding of (\ref{basic-SG}), let us first put $\Delta=0$.
The simplified Hamiltonian is
\begin{equation}
\label{interacting-hamiltonian}
H=\int \frac{dx}{2\pi} \left[uK(\pi\Pi)^2
+\frac{u}{K}(\partial_{x}\phi)^2\right]
\end{equation}

Computing the correlation functions at 0K for (\ref{interacting-hamiltonian}) in 
Matsubara time gives~:
\begin{eqnarray}
\langle T_\tau e^{\imath n \phi(x,\tau)}e^{-\imath n
\phi(0,0)}\rangle& =
&\left(\frac{x^2+(u\tau)^2}{\alpha^2}\right)^{-\frac{n^2K}{2}}
\label{phi-phi} \\
\langle T_\tau e^{\imath n \theta(x,\tau)}e^{-\imath n
\theta(0,0)}\rangle & = &
\left(\frac{x^2+(u\tau)^2}{\alpha^2}\right)^{-\frac{n^2}{2K}}
\label{theta-theta}
\end{eqnarray}
 Thus  $K$ controls the power law decay of the correlation functions
i.~e. the scaling dimensions of the operators.
More precisely (\ref{phi-phi}),(\ref{theta-theta}) imply that
$e^{\imath n \phi}$ has dimension $\frac{n^2 K}{2}$ and that
$e^{\imath n \theta}$ has dimension $\frac{n^2 }{2K}$.
$u$ can be interpreted as the velocity of the excitations.
To compute the spin-spin correlation functions, one uses the
expressions (\ref{bosonized-spin}).

For $\Delta \neq 0$,   correlation
functions cannot be obtained exactly anymore. However, the sine-Gordon 
Hamiltonian 
can still be studied using renormalization group (RG) techniques
 \cite{emery_revue_1d,giamarchi_logs}.
The flow equations for $K$ and $\Delta$ are of the Kosterlitz-Thouless form
\cite{kosterlitz_renormalisation_xy,jose_planar_2d}. From (\ref{phi-phi}), 
$\Delta$ has scaling dimension $2-\beta^2 K/4$. 
A small $\Delta$
is thus irrelevant for $K>K_c=8/\beta^2$.

When $\Delta$ is irrelevant, $K$ flows to a fixed point value $K^*$
and correlation functions keep their power law character  up to
logarithms \cite{giamarchi_logs} with the bare value of $K$ replaced
by $K^*$.
 On the other hand if $\Delta$ is relevant, $\phi$  acquires an
 expectation value that minimizes the ground state energy and a gap
is  formed.
It can then be shown\cite{emery_revue_1d} that there
 $\langle f(\phi)\rangle \sim f(\langle\phi\rangle)$
and that
$\langle T_{\tau}e^{\imath \alpha \theta(x,\tau)}e^{-\imath \alpha 
\theta(0,0)}\rangle
 \sim \exp(-\frac{\sqrt{x^2+(u\tau)^2}}{\xi})$ where $\xi$ is a
correlation length.
These results are used extensively in the paper.

\section{Calculation of the VBS order parameter in the phase
identified as a Haldane gap phase} \label{VBS-bosonization}

In this appendix  the identification of sector II of
table \ref{table3} with a Haldane gap phase is made more precise.
 It is a well known fact
that a Haldane gap phase has a hidden topological long range order
\cite{nijs_dof,tasaki_spin1} . The order parameter that measures the hidden 
topological
order\cite{nijs_dof} is known as the Valence Bond Solid (VBS) order parameter.

The VBS order parameter ${\mathcal C}$  is a nonlocal order parameter
defined as
\begin{equation}
\label{VBS}
{\mathcal C}=\lim_{\mid i-j \mid \to \infty} \langle S_i^z \exp(\imath
\pi \sum_{i<n<j} S_n^z) S_j^z \rangle
\end{equation}
In the Haldane gap phase, all the spin-spin correlation functions
decay exponentially but ${\mathcal C}\neq 0$.
Physically, the fact that the VBS order parameter is non zero indicate
that if all the sites with $S^z=0$ are  removed from a spin-1 antiferromagnetic 
chain 

the remaining (``squeezed'') chain has antiferromagnetic
order.

In the following, it is shown using bosonization that the mean value
of the VBS order parameter is non zero in the phase we identified as a
Haldane gap phase. A related problem is the dimerized spin chain
\cite{hida_vbs}, where a bosonization transformation permits to show
explicitly the existence of hidden long range order.

In the first place, a bosonized expression of the string operator $\exp(\imath 
\pi 
\sum_{i<n<j}(S_1^z+S_2^z))$ has
to be obtained.
A naive bosonization would lead  to a string operator of the form:
\begin{equation}
\exp(\imath\sqrt{2}(\phi_s(ja)-\phi_s(ia)+\int_{ia}^{ja} dx e^{\imath
\pi\frac{x}{a}}\frac{\cos(\sqrt{2}\phi_s)\cos(\sqrt{2}\phi_a)}{\pi\alpha}))
\end{equation}
That expression has very bad features.
First, in the antiferromagnetic phase, the oscillating term is non
zero and creates a complicated variable phase, that is extremely
tricky to handle because of the unknown lattice renormalizations . In
particular it is impossible with that expression to get the mean value
of the VBS order parameter.
Moreover, in the Haldane phase  the mean value of the oscillating term is zero
but if it is dropped of the calculation of the VBS order
parameter , an  unphysical zero VBS order parameter results. However, 
these incorrect results are  only due to an inappropriate boson representation 
of 
the string
order parameter.
The good representation is obtained using the identity:
$\exp(\imath \pi (S_1^z+S_2^z))=-\exp(\imath \pi(S_1^z-S_2^z))$ .This
operator identity comes from the fact that $S_z= \pm1/2$ so that the
string order parameter can now be rewritten
\begin{equation}
\prod_{i<n<j}\exp(\imath\pi(S_1^z+S_2^z))=
(-)^{j-i-1}\exp(\sum_{i<n<j}\imath\pi(S_1^z-S_2^z))
\end{equation}
The new form of the string order parameter can be straightforwardly bosonized
in the form $(-)^{i-j} e^{\imath \sqrt{2}(\phi_a(ia)-\phi_a(ja))}$.
This one has the correct sign alternation in the antiferromagnetic phase.
Using the bosonized expression of $S_1^z+S_2^z=S^z$
\begin{equation}
 S^z(x)=\frac{-\sqrt{2}\partial_x\phi_s}{\pi}+e^{\imath\pi
x/a}\frac{2\cos(\sqrt{2}\phi_s)\cos(\sqrt{2}\phi_a)}{\pi\alpha}
\end{equation}
 The VBS order parameter is obtained in the form
\begin{equation}
{\mathcal C}=\lim_{\mid x-y \mid \to \infty} \langle\cos
(\sqrt{2}\phi_s(x)) \cos(\sqrt{2}\phi_s(y))\rangle
\end{equation}
Since $\phi_s$ is long range ordered, a non zero VBS order  
parameter is obtained in the phase  identified as a Haldane gap phase
in section \ref{2chains-ferro}.
More precisely, ${\mathcal C}\sim (\langle\cos (\sqrt{2}\phi_s
)\rangle)^2$.
An alternative derivation can be found in Ref. \onlinecite{shelton_spin_ladders}.

\section{Derivation of the bosonized coupling for a random z exchange}
\label{remplissage}
For the sake of simplicity,  the derivation is made for one chain.

It is done in two steps.
First, one goes to the continuum limit
\begin{equation}
H_{\text{random Jz}}=\int dx J_z(x) S^z(x) S^z(x+a)
\end{equation}
with $J_z(x)=a J_n$,$x=na$,$S^z(x)=S_n/a$.

Then, one fermionizes using the Jordan Wigner transformation 
\ref{Jordan-Wigner}.

After straightforward calculations the $2k_F$ part of
$S^z(x) S^z(x+a)$ is obtained in the form
\begin{equation}
-(\psi^\dagger_R\psi_R+\psi^\dagger_L\psi_L)(x)(\psi^\dagger_R\psi_L
+\psi^\dagger_L\psi_R)(x+a)
+(\psi^\dagger_R\psi_L+\psi^\dagger_L\psi_R)(x)(\psi^\dagger_R\psi_R+
\psi^\dagger_L\psi_L)(x+a)
\end{equation}
Written as above, this is a purely formal expression that contains
hidden infinities.
For it to make sense, it must be normal ordered .
note that in the preceding cases,  only    two fermion
operators had to be normal ordered . Thus, normal ordering gave only some 
constants that  could
be safely dropped from the Hamiltonian.
Here, a product of 4 fermion operators has to be normal ordered. Thus,
contractions
can leave us with generated 2 particle operators that cannot be
obtained in a naive bosonization procedure. \\
Technically, Wick theorem says that any product of operators can be
written as a
sum of normal ordered products with pairings \cite{bogolyubov_ITQF}.
The two non zero pairings are the following
\begin{eqnarray}
\underline{\psi_R(x)\psi^\dagger_R(x')}=\frac{1}{2\pi(x-x')} \nonumber \\
\underline{\psi_L(x)\psi^\dagger_L(x')}=\frac{-1}{2\pi(x-x')} \nonumber \\
\end{eqnarray}

Let us pick up the first term with a non-zero contraction
\begin{eqnarray}\label{formal-coupling}
\psi^\dagger_R(x)\psi_R(x)\psi^\dagger_R(x+a)\psi_L(x+a) & = & 
:\psi^\dagger_R(x)\psi_R(x)\psi^\dagger_R(x+a)\psi_L(x+a): \nonumber \\
 &   & +:\psi^\dagger_R(x)
\underline{\psi_R(x)\psi^\dagger_R(x+a)}\psi_L(x+a): \nonumber \\
 & = & 
:\psi^\dagger_R(x)\psi_R(x)\psi^\dagger_R(x+a)\psi_L(x+a):+
\frac{-1}{2\imath\pi  a}:\psi^\dagger_R(x)\psi_L(x+a):
\end{eqnarray}
i.~e. upon normal ordering, the short distance expansions of operator
products generates a 2 fermion term that cannot be obtained doing a
naive bosonization.

Looking carefully at  expression (\ref{formal-coupling}), one can see that
it contains 4 generated 2 fermion terms plus the fully normal ordered
4 fermion terms.

Collecting together the 4 generated 2 fermion terms, one gets
\begin{equation}
\frac{1}{2\imath \pi
a}\left[:\psi^\dagger_R(x)\psi_L(x+a):-:\psi_L(x)\psi_R(x+a):+
:\psi^\dagger_R(x)\psi_L(x+a):-:\psi_L^\dagger(x)\psi_R(x+a)\right]
\end{equation}
Upon bosonization, this gives
\begin{equation}
\frac{1}{\imath \pi a}\times\frac{1}{2\pi a}\left(e^{\imath
2\phi}-e^{-\imath 2\phi}\right)=\frac{\sin 2\phi}{(\pi a)^2}
\end{equation}
Thus, one obtains
\begin{equation}\label{rdmjz}
H_{\text{random Jz}}=\int dx \frac{J^{2k_F}_z(x)}{(\pi a)^2}\sin
2\phi(x) +\text{less relevant terms \ldots}
\end{equation}

The normal ordered 4 fermion terms reduces to a term
$(\partial_x\phi)^2\sin 2 \phi$. Power counting
implies that that
term has dimension $2+K$ and is thus irrelevant irrespective of the
value of $K$.

It is possible to check that (\ref{rdmjz}) is the correct expression
by the following argument: $S^z_iS^z_{i+1}$ is invariant under a
rotation around the z axis, so that it cannot depend on $\theta$
 and since it must have the same
scaling dimension at the isotropic point as $S^x_i S^x_{i+1}$, its bosonized 
form 
must be
either $\cos 2\phi$  or $\sin 2\phi$.
Since  $S^z \to -S^z$   corresponds  in the
bosonization language to
$\phi \to \frac{\pi}{2}-\phi$ the bosonized form  of  $S^z_i S^z_{i+1}$
must be invariant under such transformation. This rules out the $\cos
2\phi$ and leads to (\ref{rdmjz}) as the only possible expression for
the bosonized form of $S_i^zS_{i+1}^z$. However, it is comforting to
be able to derive directly the expression (\ref{rdmjz}), since it
proves that (\ref{rdmjz}) is not an ad hoc expression that is chosen
arbitrarily in order
to obtain physically sound results from an inappropriate technique.

\bibliographystyle{prsty}
%\bibliography{totphys,spinchains}

\begin{table}
\caption{the 4 sectors of the pure 2 spin chains  model with $J_{\perp}<0$}
\begin{tabular}{ccccc}
 &I&II&III&IV\\
\tableline
$K_s$&$<1$&$<1$&$>1$&$>1$ \\
$K_a$&$<1/2$&$>1/2$&$<1/2$&$>1/2$ \\
\tableline
$\phi_s$&$\langle\phi_s\rangle =0$&$\langle\phi_s\rangle
=0$&massless&massless\\
$\theta_a,\phi_a$&$\langle\phi_a\rangle =0$&$\langle\theta_a\rangle
=0$&$\langle\phi_a\rangle =0$&$\langle\theta_a\rangle =0$\\
phase & Ising AF & Haldane gap & $XY2$ & $XY1$\\
Order Parameter&$\cos(\sqrt{2}\phi_s)\cos(\sqrt{2}\phi_a)$&?&
$e^{\imath\sqrt{2}\theta_s}$&$e^{\imath\frac{\theta_s}{\sqrt{2}}}
\cos(\frac{\theta_a}{\sqrt{2}})$
\end{tabular}
\label{table3}
\end{table}

\begin{table}
\caption{the 4 sectors of the pure 2 spin chains  model with $J_{\perp}>0$}
\begin{tabular}{ccccc}
 &I&II&III&IV\\
\tableline
$K_s$&$<1$&$<1$&$>1$&$>1$ \\
$K_a$&$<1/2$&$>1/2$&$<1/2$&$>1/2$ \\
\tableline
$\phi_s$&$\langle\phi_s\rangle
=\frac{\pi}{\sqrt{8}}$&$\langle\phi_s\rangle
=\frac{\pi}{\sqrt{8}}$&massless&massless\\
$\theta_a,\phi_a$&$\langle\phi_a\rangle
=\frac{\pi}{\sqrt{8}}$&$\langle\theta_a\rangle
=\frac{\pi}{\sqrt{2}}$&$\langle\phi_a\rangle
=\frac{\pi}{\sqrt{8}}$&$\langle\theta_a\rangle
=\frac{\pi}{\sqrt{2}}$\\
phase & Ising AF & singlet & $XY2$ & $XY1$\\
Order Parameter&$\sin(\sqrt{2}\phi_s)\sin(\sqrt{2}\phi_a)$&VBS&
$e^{\imath\sqrt{2}\theta_s}$&$e^{\imath\frac{\theta_s}{\sqrt{2}}}
\sin(\frac{\theta_a}{\sqrt{2}})$
\end{tabular}
\label{table2}
\end{table}

\end{document}